\documentclass[11pt,a4paper]{article}
\usepackage[english]{babel}
\usepackage{amsmath,amsthm,amssymb,epsfig,latexsym}
\usepackage{color}

\newcommand{\gr}{\color{green}}

\usepackage{ulem}

\newcommand{\mb}[1]{\quad\mbox{#1}\quad}

\newcommand{\so}{\scriptscriptstyle \rm I}
\newcommand{\st}{\scriptscriptstyle \rm I\hspace{-1pt}I}
\newcommand{\qo}{\rm i}
\newcommand{\qt}{\rm ii}
\newcommand{\be}[1]{\begin{equation}\label{#1}}
\newcommand{\ba}[1]{\begin{multline}\label{#1}}
\newcommand{\ee}{\end{equation}}
\newcommand{\ea}{\end{eqnarray}}

\newcommand{\str}{\mathop{\rm str}}

\newtheorem{prop}{Proposition}[section]
\newtheorem{lemma}{Lemma}[section]

\def\nonu{\nonumber\\ }
\newcommand{\bea}{\begin{eqnarray}}
\newcommand{\eea}{\end{eqnarray}}
\def\qed{\hfill\nobreak\hbox{$\blacksquare$}\par\medbreak}
\newcommand{\prof}{{\textsl{Proof:}\ }}

\def\fe{{\mathfrak e}}

        \def\cR{{\cal R}}
        \def\cU{{\cal U}}
    \def\cW{{\cal W}}

\newcommand{\CC}{{\mathbb C}}
\newcommand{\EE}{{\mathbb E}}
\newcommand{\II}{{\mathbb I}}

\newcommand{\RR}{{\mathbb R}}
\newcommand{\TT}{{\mathbb T}}

\newcommand{\ZZ}{{\mathbb Z}}

\newcommand{\vph}{\varphi}

\newcommand{\wt}[1]{\widetilde{#1}}
\newcommand{\wb}[1]{\overline{#1}}

 \makeatletter
 \@addtoreset{equation}{section}
 \makeatother
 
\def\bu{{\bar u}}   \def\bv{{\bar v}}
\def\gr{\mathbf{gr}}
\def\EE{{\rm E}}

\begin{document}

\begin{flushright}
LAPTH-015/16
\end{flushright}

\vspace{22pt}

\begin{center}
\begin{LARGE}
{\bf Bethe vectors for models based on\\[2mm] the super-Yangian $Y(\mathfrak{gl}(m|n))$}
\end{LARGE}

\vspace{40pt}

\begin{large}
S.~Z.~Pakuliak${}^{a,b}$, E.~Ragoucy${}^c$, N.~A.~Slavnov${}^d$\  \footnote{
 stanislav.pakuliak@jinr.ru, eric.ragoucy@lapth.cnrs.fr, nslavnov@mi.ras.ru}
\end{large}

 \vspace{12mm}

${}^a$ {\it Moscow Institute of Physics and Technology,  Dolgoprudny, Moscow reg., Russia}

\vspace{4mm}

${}^b$ {\it Laboratory of Theoretical Physics, JINR,  Dubna, Moscow reg., Russia}

\vspace{4mm}

${}^c$ {\it Laboratoire de Physique Th\'eorique LAPTH, CNRS and Universit\'e de Savoie,\\
BP 110, 74941 Annecy-le-Vieux Cedex, France}

\vspace{4mm}

${}^d$ {\it Steklov Mathematical Institute, Moscow, Russia}

\end{center}


\vspace{4mm}


\begin{abstract}
\noindent We study Bethe vectors of integrable models based on the super-Yangian $Y(\mathfrak{gl}(m|n))$.
Starting from the super-trace formula, we exhibit recursion relations for these vectors in the case of $Y(\mathfrak{gl}(2|1))$ and $Y(\mathfrak{gl}(1|2))$.
These recursion relations allow to get explicit expressions for the Bethe vectors.
Using an antimorphism of the super-Yangian $Y(\mathfrak{gl}(m|n))$, we also construct a super-trace formula for dual Bethe vectors, and, for $Y(\mathfrak{gl}(2|1))$ and $Y(\mathfrak{gl}(1|2))$ super-Yangians, show recursion relations for them.
Again, the latter allow us to get explicit expressions for dual Bethe vectors.
\end{abstract}


\setcounter{footnote}{0}

\section{Introduction}
Algebraic Bethe Ansatz (ABA) is a powerful tool for the calculation of Bethe vectors (BVs) of integrable models, which allows to access to the correlation functions of these models, through the calculation of the form factors (see e.g. \cite{FadST79,BogIK93L,FadLH96,IZ1984} and references therein). It has been successfully applied for models based on $\mathfrak{gl}(2)$ or its quantum deformation. In that case, one uses a determinant presentation for the 
norm and the scalar product of BVs \cite{IzKo,Nikita1} to get an easy-to-handle expression for the form factors. The latter then can be used to study the
thermodynamic limit of the form factors and get insight on the correlation functions asymptotic behavior \cite{asympt,asympt2}. 

However for models based on higher rank Lie algebras (or their quantum deformation), much less is known. Already at the level of BVs,
although the ABA was performed long ago \cite{KR1982}, few explicit expressions have been obtained, apart from the scalar product obtained in \cite{Res86} that is difficult to handle. Recently, in a series of papers, the case of $\mathfrak{gl}(3)$ and of its quantum deformation was successfully studied, starting from explicit forms for BVs \cite{BVs,BelPRS13d} and the calculation of their scalar products \cite{PRStrigo,Nikita2}, up to determinant presentations for form factors \cite{FF}. The case of Bose gas with two internal degrees of freedom was also tackled, again with explicit expressions for Bethe vectors \cite{compo1,nikita}, and determinant presentations of the form factors \cite{PozOK12,compo2,bose}. The case of more general Lie algebra (or their quantum deformation) remains to be done, but some steps have been done towards their resolution, using the current presentation of these algebras \cite{KhEOPS}. Note also that a trace formula for BVs is known for the $\mathfrak{gl}(n)$ \cite{TV} that can be used to deduce more properties of Bethe vectors.

The case of superalgebras is even less rich, apart from a super-trace formula in the $\mathfrak{gl}(m|n)$ case \cite{BR1}.
 It is rather unfortunate, given their
 relevance in the study of gauge theories \cite{FW,QCD,assoc}, in particular super-Yang-Mills (SYM) theories and AdS/CFT correspondence (see \cite{AdS} and references therein). Indeed it is now believed that integrability should play an important role in SYM theories based on $PSL(4|4)$ \cite{SYMgl44}, and also in its subsectors, such as $PSL(2|2)$ or $SL(2|1)$ \cite{SYMgl22}. Moreover, the t-J model, well-known in condensed matter physics, is based on the $\mathfrak{gl}(2|1)$ superalgebra \cite{tJ}.
 Thus, there is some urge to find explicit representations for the Bethe vectors for integrable models based on these superalgebras. The aim of this paper is to present explicit expressions for Bethe vectors of integrable models based on $\mathfrak{gl}(2|1)$ and $\mathfrak{gl}(1|2)$.

The method we will be using mimics the one used for the $\mathfrak{gl}(3)$. Starting from the super-trace formula, we will deduce some recursion relations obeyed by the BVs. Then, solving these recursions, we will obtain explicit expressions for BVs. Using different morphisms, we will use the solution for the $Y(\mathfrak{gl}(2|1))$ case to construct solutions for the $Y(\mathfrak{gl}(1|2))$ case and also for the dual BVs.

The plan of our paper follows the lines we mentioned. After reminding some properties of Yangians $Y(\mathfrak{gl}(m|n))$, based on $\mathfrak{gl}(m|n)$ Lie superalgebras in section \ref{sect:glmn}, we will remind the super-trace formula for the case of $Y(\mathfrak{gl}(2|1))$ and $Y(\mathfrak{gl}(1|2))$ in section \ref{sect:supertr}.
Then, using the super-trace formula, we will show in section \ref{sect:1strecur} that the BVs obey a recursion relation. A second recursion relation will be exhibited in section \ref{sect:2ndrecur}. These recursion relations allow to get explicit formulas for BVs of $Y(\mathfrak{gl}(2|1))$ models (section \ref{sect:explY21}) and then for $Y(\mathfrak{gl}(1|2))$ models and for dual BVs (section \ref{sect:expl-remain}).

\section{$\mathfrak{gl}(m|n)$  rational $R$-matrices\label{sect:glmn}}
\subsection{Graded vector spaces}
We will work with graded vector spaces. We introduce the  $\mathbb{Z}_2$-grading
$$
[\cdot]\ :\  \{1,2,...,m+n\} \ \to\ \{0,1\},
$$
where $[j]=0$ for $m$ of the above integers, while $[j]=1$ for the remaining $n$ integers.
The basic vector space will be $\CC^{m+n}$, equipped with a gradation that we will loosely write $[\cdot]$:
$$
[\cdot]: \left\{\begin{aligned}  &\CC^{m+n} &\to\quad &\ZZ_2 \\ &e_a  &\to\quad &[a],\end{aligned}\right.
$$
so that the vector space will be noted $\CC^{m|n}$.  The elementary matrices of End$(\CC^{m|n})$ will be graded accordingly
$[\EE_{ij}]=[i]+[j]$. The grading is a morphism for the multiplication, so that
$$[\EE_{ij}\EE_{kl}]=[\EE_{ij}]+ [\EE_{kl}]=[i]+[j]+[k]+[l].$$
Vectors or matrices that have $\ZZ_2$-grading 0 (resp. 1) will be called even (resp. odd).

The  gradation we will be mainly using has the form
\be{disting-grad}
[i]=\begin{cases} 0\,,\quad i=1,2,...m\,,\\
1\,,\quad i=m+1,...,m+n\,.
\end{cases}
\ee
We will call it the \textsl{distinguished gradation}, because it corresponds to the grading associated to the distinguished Dynkin diagram of $\mathfrak{gl}(m|n)$, with only one fermionic simple root. Other gradations can be defined, such as
\be{ferm-grad}
\begin{cases} [2i-1]=1\,,\\ [2i] =0\,,\end{cases} \quad i=1,2,...
\ee
which leads to  a 'grey' Dynkin diagram where the   simple roots are all fermionic. A third example is given by
\be{alt-grad}
\begin{cases} [4i+1]=[4i+2]=1\,,\\ [4i+3]=[4i+4] =0\,,\end{cases} \quad i=0,1,2,...
\ee
which leads to an 'alternating' Dynkin diagram, where the simple roots are alternatively fer\-mio\-nic and bosonic.
Hereafter, by $[\cdot]$ we will always understand the distinguished gradation given by \eqref{disting-grad}.

On End$(\CC^{m|n})$ we define a super-trace operator which is graded-cyclic:
$$
{\str}(\EE_{ij})=(-1)^{[j]}\,\delta_{ij}\,,\qquad {\str}(\EE_{ij}\,\EE_{kl})=(-1)^{([i]+[j])([k]+[l])}\,{\str}(\EE_{kl}\,\EE_{ij}).
$$
We also define a supertransposition
\be{supertranspo}
\EE_{ij}^t=(-1)^{[j][i]+[j]}\,\EE_{ji}\,.
\ee
The graded transposition is compatible with the super-trace: for any matrices $A$ and $B$
\be{transpo-trace}
{\str} A = {\str} A^t \mb{and} {\str}(A^t\,B^t)={\str}(A\,B).
\ee
Note that the graded transposition is idempotent\footnote{One could define
a transposition $(\EE^T)_{ij}=(-1)^{[i][j]} \EE_{ji}$, which is an antimorphism of order 2. However it does not obey relation \eqref{transpo-trace}.} of order 4, \underline{not}  of order 2, a common feature in superalgebras. Indeed for a matrix $A$ of given grade, we have
\be{transpo-squared}
 \big(A^t\big)^t = (-1)^{[A]}\, A.
\ee

Tensor products of $\CC^{m|n}$ spaces will be also graded:
$$
(\II\otimes \EE_{ij})\,\cdot\,(\EE_{kl}\otimes\II) = (-1)^{([i]+[j])([k]+[l])}\,\EE_{kl}\otimes \EE_{ij}.
$$
 where $\II=\sum_{i=1}^{m+n}\EE_{ii}$ is a unit matrix.

Together with this grading of spaces and tensor products, one defines a graded permutation operator
$$
P=\sum_{i,j=1}^{m+n} (-1)^{[j]}\, \EE_{ij}\otimes \EE_{ji}
$$
which obeys
$$
P^2=\II\otimes\II\ \mb{and}
P^{t_1t_2}=P.
$$
It acts on tensor products of vectors and matrices as:
$$
 P\,e_i\otimes e_j = (-1)^{[i][j]}\, e_j\otimes e_i \mb{and}
P\,\EE_{ij}\otimes \EE_{kl}\,P = (-1)^{([i]+[j])([k]+[l])}\, \EE_{kl}\otimes \EE_{ij}.
$$

\subsection{Super-Yangian algebras $Y\big(\mathfrak{gl}(m|n)\big)\equiv Y(m|n)$}
The super-Yangian $Y\big(\mathfrak{gl}(m|n)\big)\equiv Y(m|n)$ is
 an associative algebra with a unit element $\mathbf{1}$ generated by the modes
$T_{ij}^{(p)}$, $p\in\ZZ_+$
of the
{\it universal monodromy matrix}
 \be{series:T}
 T(u)=\sum_{i,j=1}^{m+n} \EE_{ij}\otimes T_{ij}(u)=
 \II\otimes \mathbf{1}+\sum_{i,j=1}^{m+n}\sum_{p=0}^\infty \EE_{ij}\otimes T_{ij}^{(p)}\,\left(\frac{c}{u}\right)^{p+1},
 \ee
defined by the $RTT$ relation:
\bea
&&R_{12}(u_1,u_2)\, T_1(u_1)\, T_2(u_2) =  T_2(u_2)\, T_1(u_1)\,R_{12}(u_1,u_2)\label{rtt}\\
&&\mb{with} R_{12}(u_1,u_2) = \II +g(u_1,u_2)\,P_{12}
\mb{and} g(u,v)=\frac{c}{u-v},\label{rmat}
\eea
where $c$ is  a constant. We will loosely say that the series
 $T_{ij}(u)$ belongs to the super-Yangian, although strictly speaking they belong to $Y(m|n)[[u^{-1}]]$.
The relation \eqref{rtt} is written in the tensor product
${\rm End}(\CC^{m|n})\otimes  {\rm End}(\CC^{m|n})\otimes Y(m|n)[[u^{-1}]]$, and
the indices indicate which copy of End$(\CC^{m|n})$ the operators belong to.

 The $\mathbb{Z}_2$-grading $[\cdot]$ is extended to the super-Yangian through
 $$
 [T_{ij}(u)]=[T_{ij}^{(p)}]=[i]+[j]\,,\quad \forall u\in\CC\,,\quad \forall p\in\ZZ_+\,.
 $$
As for matrices, generators that have $\ZZ_2$-grading 0 (resp. 1) will be called even (resp. odd).
 Relations between these generators are given below, but let us first stress
 that, because of the graded tensor product,  the order in the tensor product matters. Indeed, for instance $T(u)\,\EE_{kl}=\sum_{i=1}^{m+n}(-1)^{([i]+[k])([k]+[l])}\,\EE_{il}\otimes T_{ik}(u)$ while for $\wt T(u)=\sum_{i,j=1}^{m+n} T_{ij}(u)\otimes \EE_{ij}$, one has
 $\wt T(u)\,\EE_{kl}=\sum_{i=1}^{m+n} T_{ik}(u)\otimes \EE_{il}$.

The $R$-matrix \eqref{rmat} is unitary  and symmetric:
\bea
&&R_{21}(u_2,u_1)\,R_{12}(u_1,u_2)=f(u_1,u_2)\,f(u_2,u_1)\,\II\otimes\II
\,,\qquad\\
&&R_{21}(u_1,u_2)=P\,R_{12}(u_1,u_2)\,P=R_{12}(u_1,u_2)=R_{12}(u_1,u_2)^{t_1t_2}\,,
\eea
 where  $f(u,v)=1+g(u,v)$. 

Projecting the relation \eqref{rtt}, one gets the commutation relations for the super-Yangian $Y(m|n)$:
\be{comT-1}
{[T_{ij}(z)\,,\,T_{kl}(w)\}} = (-1)^{[l]([i]+[j])+[i][j]}\,g(z,w)\Big(T_{il}(z)\,T_{kj}(w) - \,T_{il}(w)\,T_{kj}(z)\Big) ,
\ee
where we have introduced the graded commutator
$$
{[T_{ij}(z)\,,\,T_{kl}(w)\}} =T_{ij}(z)\,T_{kl}(w) - (-1)^{([i]+[j])([k]+[l])}\, T_{kl}(w)\,T_{ij}(z).
$$
 When $n=0$ we recover the  Yangian based on $\mathfrak{gl}(m)$.

Remark that the monodromy matrix $T(u)$ is globally even, because the $\ZZ_2$-grading of $\EE_{ij}$ is the same as the one of $T_{ij}(u)$. The same is true for the $R$-matrix.

Note that, by definition, the commutator is graded anti-symmetric:
\be{}
{[T_{ij}(z)\,,\,T_{kl}(w)\}} = -(-1)^{([i]+[j])([k]+[l])}\,{[T_{kl}(w)\,,\,T_{ij}(z)\}},
\ee
which implies in particular that
\bea
{[T_{ij}(z)\,,\,T_{kl}(w)\}} &=& (-1)^{[i]([k]+[l])+[k][l]}\,g(z,w)\Big(T_{kj}(w)\,T_{il}(z) - \,T_{kj}(z)\,T_{il}(w)\Big).
\label{comT-2}
\eea

%
The graded transfer matrix is defined as
\be{transfer.mat}
t(w)=\str T(w)= \sum_{j=1}^{m+n} (-1)^{[j]}\, T_{jj}(w).
\ee
It defines an integrable system, due to the relation $[t(z)\,,\,t(w)]=0$.

\subsubsection*{Evaluation map and  subalgebras of $Y(m|n)$}
The generators $T^{(0)}_{i,j}$, $i,j=1,2,...,m+n$, form a Lie superalgebra $\mathfrak{gl}(m|n)$ with commutation relations
$$
{[T_{ij}^{(0)}\,,\,T_{kl}^{(0)}\}} = (-1)^{[i]([k]+[l])+[k][l]}\Big(T_{kj}^{(0)}\,\delta_{il} - \,\delta_{kj}\,T_{il}^{(0)}\Big).
$$
They act naturally on the monodromy matrix:
$$
{[T_{ij}^{(0)}\,,\,T_{kl}(z)\}} = (-1)^{[i]([k]+[l])+[k][l]}\Big(T_{kj}(z)\,\delta_{il} - \,\delta_{kj}\,T_{il}(z)\Big).
$$

In fact, as for the Yangian $Y(\mathfrak{gl}(m))$, there exists an {\it evaluation} morphism from $Y(m|n)$ to the Lie superalgebra $\mathfrak{gl}(m|n)$:
$$
ev:\begin{cases}Y(m|n) \ \to\ \mathfrak{gl}(m|n)\\
\displaystyle{T(u)\ \to\ \II \otimes \mathbf{1}+\frac{c}{u} \sum_{i,j=1}^{m+n} \EE_{ij}\otimes \fe_{ij}}
\end{cases}\mb{where} \quad \fe_{ij}\in gl(m|n),
$$
where the $\mathfrak{gl}(m|n)$ generators $\fe_{ij}$ just corresponds to the so-called \textit{zero modes}
$T^{(0)}_{ij}$. The latter are a symmetry of the integrable model described by the transfer matrix:
$[T^{(0)}_{ij}\,,\, t(z)]=0.$

As far as subalgebras are concerned, the generators $T_{ij}(u)$, $1\leq i,j\leq m$ (resp. $m+1\leq i,j\leq m+n$) form a Yangian $Y(\mathfrak{gl}(m))$ (resp. $Y(\mathfrak{gl}(n))$) subalgebra of $Y(m|n)$. However, they are not Hopf-subalgebras.

\subsubsection*{Shorthand notation used in the paper}
To make the presentation easier to read, we will use the following notation throughout the paper.

Sets of parameters will be noted with a bar, such as $\bar u=\{u_1,u_2,...,u_\ell\}$. Let us stress that these sets will be ordered, due to the $\ZZ_2$-grading. Elements of these sets will have a latin index, e.g. $u_j$, while subsets will have, as a rule, a roman index, e.g. $\bar u_{\so}$.
These subsets will come as partitions of the original set, so that when saying $\bar u$ is divided into $\bar u_{\so}$ and $\bar u_{\st}$, it will mean $\bar u_{\so}\cap\bar u_{\st}=\emptyset$ and $\bar u_{\so}\cup\bar u_{\st}=\bar u$. There will be one exception to these rules:
the subset $\bar u_j=\bar u\setminus \{u_j\}$.

When considering functions of one or two variables, as $\lambda(z)$ or $f(z,w)$, the notation $\lambda(\bar u)$ will mean product of the $\lambda(u_j)$ for each element $u_j$ in the set $\bar u$. In the same way, $f(\bar u_{\st},z)$ will mean product of the $f(u_j,z)$ for each element $u_j$ in the subset $\bar u_{\st}$, and $f(\bar u_{\st},\bar v)$ will mean the double product of $f(u_j,v_k)$ factors for each element $u_j$ in the subset $\bar u_{\st}$ and each element in the set $\bar v$. These rule will also apply to operators that commute at different values of the parameters.

\subsection{Representations and morphisms of super-Yangians\label{sect:morphism}}

\subsubsection*{Automorphisms of $Y(m|n)$}
The $R$-matrix and the monodromy matrix being globally even, it is easy to show from \eqref{rtt} that
\be{psi}
\psi: \begin{array}{l} T(u) \ \to\ T^t(u) \\ T_{ij}(u)\ \to\ (-1)^{[i]\,[j]+[i]}\,T_{ji}(u)\end{array}
\mb{is an antimorphism of $Y(m|n)$.}
\ee
 We call this antimorphism a {\it transposition map}.
Let us point out the difference of sign factor between $\EE_{ij}^t$ and $\psi(T_{ij}(u))$, see \eqref{supertranspo} and \eqref{psi}, which ensures that $\psi(T(u))=T^t(u)$.
One can also check directly that the commutation relations \eqref{comT-1} are consistent with the relations (for $A$ and $B$ of definite grading)
$$
\psi\big([A\,,\,B]\big) = - \big[\psi(A)\,,\, \psi(B)\big] \mb{and} \psi(AB)= (-1)^{[A][B]}\, \psi(B)\psi(A).
$$
As a byproduct, \eqref{transpo-squared} shows that the application $\gr$ which multiplies any generator by its $\ZZ_2$-grade
\be{psi2}
\gr\,:\ T_{ij}(u)\ \to\ (-1)^{[i]+[j]}\, T_{ij}(u)
\ee
 is an automorphism of the algebra, since $\gr=\psi\,\circ\,\psi$.  Another way to see that $\gr$ is an automorphism is to realize that it corresponds to a conjugation
 $$
{\rm Ad}_\omega : T(u)\ \to\ \omega\,T(u)\,\omega^{-1}\mb{with}\omega=\sum_{k=1}^{m+n} (-1)^{[k]}\,\EE_{kk}.
 $$

To construct Bethe vectors we will consider the right  representation of the super-Yangian
$Y(m|n)$ generated by a {\it singular} vector $\Omega$ such that
\be{right-rep}
T_{ii}(u)\Omega=\lambda_i(u)\Omega\,,\quad T_{ij}(u)\Omega=0\,,\quad\mbox{for}\quad i>j\,.
\ee
For the dual Bethe vectors we will apply the transposition map $\psi$ \eqref{psi} to obtain the left representations of super-Yangian
$Y(m|n)$ generated by the vector $\Omega^\dag=\psi(\Omega)$ such that
\be{left-rep}
\Omega^\dag T_{ii}(u)=\lambda_i(u)\Omega^\dag \,,\quad \Omega^\dag
T_{ij}(u)=0\,,\quad\mbox{for}\quad i<j\,.
\ee
The transposition maps right highest weight representations into left lowest weight representations. In that case, the weights
of the left- and right- representations will be the same.

\subsubsection*{Isomorphism between $Y(m|n)$ and $Y(n|m)$}
We consider now a morphism between $Y(m|n)$ and $Y(n|m)$.
\begin{prop}\label{prop.iso}
Let $T_{ij}(u)$ be the generators of the Yangian $Y(m|n)$,
and $\wt T_{ij}(u)$ be the generators of the Yangian $Y(n|m)$. Let $[\cdot]$ (resp. $\wt{[\cdot]}$) be the $\ZZ_2$-grading of the
$Y(m|n)$ super-Yangian (resp. $Y(n|m)$ super-Yangian). Then, the following mapping:
\be{def:vph}
\vph\ :\ \begin{cases}
Y(m|n)\quad\  \to\quad Y(n|m)\\[1ex]
T_{ij}(u)\qquad  \to\quad   (-1)^{[i]\,[j]+[j]+1}\,\wt T_{\wb\jmath\,\wb\imath}(u)\\[1ex]
[j] \qquad\quad\ \to\quad \wt{[j]}=[\wb{\jmath}]+1
\end{cases}
\ee
where $\wb\jmath=m+n+1-j$, defines an isomorphism between $Y(m|n)$ and $Y(n|m)$ which is compatible with the supertrace operation.
\end{prop}
\prof
 The grading in $Y(m|n)$ is given by $[j]=0$ when $j\leq m$ and $[j]=1$ when $j>m$, which can be reformulated as
 $[\wb\jmath]=0$ when $\wb\jmath> n$ and $[\wb\jmath]=1$ when $\wb\jmath\leq n$. Up to a global shift of 1 modulo 2, $[\wb\jmath]$ in $Y(m|n)$ corresponds to
$\wt{[j]}$ in $Y(n|m)$. Then, by a direct calculation, starting from the relation \eqref{comT-1} in $Y(m|n)$ one gets through $\vph$ the relation \eqref{comT-2} for $Y(n|m)$. 

 In fact any multiplication by $(-1)^{[i]}$ and/or by $(-1)^{[j]}$ will keep the morphism property. We partially fix this freedom by demanding that the image of $str T(u)$ is $str \wt T(u)$: there remain only two possibilities, one is given in \eqref{def:vph}, the other one is $T_{ij}(u) \, \to \,   (-1)^{[i]\,[j]+[i]+1}\,\wt T_{\wb\jmath\,\wb\imath}(u)$.
\qed

The isomorphism $\vph$ induces isomorphism between representations of the two super-Yangians. In that case, the highest
weights map as
\be{vph-lambda}
\vph:\ (\lambda_1(u),...,\lambda_{m+n}(u))\ \to\ (\wt\lambda_1(u),...,\wt\lambda_{m+n}(u))
\mb{with} \wt\lambda_j(u)=-\lambda_{m+n+1-j}(u),
\ee
which just corresponds to the mapping of $T_{jj}(u)$ generators to $\wt T_{\wb\jmath\wb\jmath}(u)$ ones.

\subsubsection*{Compositions of morphisms}
The different morphisms can be composed, and we get relations among them. We focus on the morphisms $\vph$, $\psi$ and $\gr$, and
to clarify the presentation, we fix $m$ and $n$ and call $T_{ij}(u)$ (resp. $\wt T_{ij}(u)$) the elements of
$Y(m|n)$ (resp. $Y(n|m)$). In the same way, we use the following notation:
\be{diag-morph}
\begin{array}{ccc} Y(m|n)&\ \stackrel{\vph}{\longrightarrow}\ &Y(n|m) \\[1ex]
{\scriptstyle \psi}\,\updownarrow\,{\scriptstyle \gr}& & {\scriptstyle \wt{\gr}}\updownarrow{\scriptstyle \wt\psi} \\[1ex]
Y(m|n)&\ \stackrel{\wt\vph}{\longleftarrow}\ &Y(n|m)
\end{array}
\ee
In the above diagram, we remind that $\vph,\wt\vph,\gr$ and $\wt{\gr}$ are isomorphisms, while $\psi$ and $\wt\psi$ are antimorphisms, see section \ref{sect:morphism}.
Then, we have the following.
\begin{lemma}\label{vph-vphtilde}
We have the following composition rules
\bea
&&\wt\vph\,\circ\,\vph={ \rm id}
\mb{;}
\vph\,\circ\,\wt\vph=\wt{{ \rm id}}\mb{;}
\psi\,\circ\,\psi=\gr
\mb{;}
\wt\psi\,\circ\,\wt\psi=\wt{\gr}
\label{phi-phi}\\
&&\wt\psi\,\circ\,\vph=\vph\,\circ\,\psi\mb{;}
\psi\,\circ\,\wt\vph=\wt\vph\,\circ\,\wt\psi\mb{;}
\wt{\gr}\,\circ\,\vph=\vph\,\circ\,\gr\mb{;}
\gr\,\circ\,\wt\vph=\wt\vph\,\circ\,\wt{\gr}.\qquad
\label{psi-phi}
\eea
These rules imply in particular that the diagram \eqref{diag-morph} is commutative.
\end{lemma}
\prof Direct calculation, applying the definitions \eqref{psi}, \eqref{def:vph} and  \eqref{psi2} to $T_{ij}(u)$ or $\wt T_{ij}(u)$.
\qed

\section{Bethe vectors\label{sect:supertr}}
\subsection{Super-trace formula}
The generalization of the trace formula for Bethe vectors (introduced by Tarasov and Varchenko \cite{TV})
was given in \cite{BR1} for super-Yangians and quantum deformations of super-symmetric affine algebras. 
In the case of super-Yangians, and specializing
to the case of $Y(2|1)$ to simplify the presentation, it reads
\bea
\Phi_{a,b}(\bar u,\bar v) &=& \frac{(-1)^{b(b+1)/2}}{H(\bar v)}\, {\str}_{1\cdots a+b} \Big[ T_1(u_1)\cdots T_a(u_a)
T_{a+1}(v_1)\cdots T_{a+b}(v_b)\,\RR_{1\dots a+b}
\nonu
&&\qquad\qquad\qquad\qquad\qquad\times\,
\, \EE_{32}^{(a+b)}\cdots \EE_{32}^{(a+1)}\EE_{21}^{(a)}\cdots \EE_{21}^{(1)}\Big]\Omega
\label{def:Phi-str1}
\\
&=& \frac{(-1)^{b}}{H(\bar v)}\, {\str}_{1\cdots a+b} \Big[ T_1(u_1)\cdots T_a(u_a)T_{a+1}(v_1)\cdots T_{a+b}(v_b)\,\RR_{1\dots a+b}
\nonu
&&\qquad\qquad\qquad\qquad\qquad\times\,
\, \EE_{21}^{(1)}\cdots \EE_{21}^{(a)}\EE_{32}^{(a+1)}\cdots \EE_{32}^{(a+b)}\Big]\Omega\,,
\label{def:Phi-str2}\\
H(\bar v) &=& \prod_{1\leq k<j\leq b}h(v_j,v_k) \mb{with} h(u,v)=\frac{f(u,v)}{g(u,v)}=\frac{u-v+c}{c}\,.
\label{def:H}
\eea
where the super-trace ${\str}$ is taken over $(a+b)$ copies of the auxiliary space $\CC^{2|1}$, a graded version of $\CC^3$, and $\EE_{jk}^{(p)}$ is the elementary matrix $\EE_{jk}$ in the $p^{th}$ copy of the auxiliary space.
$\RR_{1\dots a+b}$ is the following product of $R$-matrices:
\bea
\RR_{1\dots a+b} &=& (R_{a+1,a}\dots R_{a+1,1})(R_{a+2,a}\dots R_{a+2,1})\cdots(R_{a+b,a}\dots R_{a+b,1})
\nonu
&=& (R_{a+1,a}\dots R_{a+b,a})(R_{a+1,a-1}\dots R_{a+b,a-1})\cdots(R_{a+1,1}\dots R_{a+b,1})
\label{def:RR}
\eea
where we abbreviated $R_{a+j,k}\equiv R_{a+j,k}(v_j,u_k)$.

Note that the indices $a,b$ in $\Phi_{a,b}(\bar u,\bar v)$ indicate that $\#\bu=a$ and $\#\bv=b$.

Let us stress that since the tensor space is graded, the order of the elementary matrices $\EE_{32}^{(p)}$ matters, as illustrated in \eqref{def:Phi-str1}-\eqref{def:Phi-str2}. In what follows, unless explicitly written, we will use the 'natural order', namely $(\EE_{32})^{\otimes b}$ stands for $\EE_{32}^{(a+1)}\cdots \EE_{32}^{(a+b)}$. The coefficient $(-1)^b/H(\bar v)$ in \eqref{def:Phi-str2} is for later convenience, see proposition \ref{prop.sym}.

$\Omega$ is the highest weight of the representation. It obeys \eqref{right-rep} with
\be{def:lambda}
 \lambda_j(z)=1+\frac{c}{z} \lambda_j^{(0)}+o(z^{-2})\,.
\ee

We give here some simple examples extracted directly from the super-trace formula:
\bea
\Phi_{a0}(\bar u,\emptyset) &=& T_{12}(u_1)\cdots T_{12}(u_a)\Omega\\
\Phi_{0,b}(\emptyset,\bar v) &=& \frac{1}{H(\bar v)}T_{23}(v_1)\cdots T_{23}(v_b)\Omega\\
\Phi_{11}(u,v) &=& \Big(T_{12}(u) T_{23}(v) + \lambda_2(v)\,g(v,u)\,T_{13}(u)\Big)\Omega\\
\Phi_{12}(u,\{v_1,v_2\}) &=& h(v_2,v_1)^{-1}\,T_{12}(u)\, T_{23}(v_1)\, T_{23}(v_2)\Omega  \\
&&+g(v_2,v_1)\,T_{13}(u)\,\Big(\lambda_2(v_1)\,g(v_1,u)\, T_{23}(v_2)-
 \lambda_2(v_2)\,g(v_2,u)\,T_{23}(v_1) \Big)\Omega\nonumber\\
\Phi_{21}(\bar u,v) &=& T_{12}(u_1)\, \Phi_{11}(u_2,v)  + \lambda_2(v)\,g(v,u_1)\,f(v,u_2)\,T_{13}(u_1)\,\Phi_{1,0}(u_2,\emptyset) .
\eea

The following property has been proven in \cite{BR1} for the super-Yangian $Y(m|n)$. Again, to simplify the presentation, we just reproduce it for the two specific cases $Y(2|1)$ and $Y(1|2)$.
\begin{prop}\label{prop.BV}
The BVs $\Phi_{a,b}(\bar u,\bar v)$, as defined  in \eqref{def:Phi-str2} for the super-Yangian $Y(2|1)$ or in \eqref{def:Phitilde-str} for the super-Yangian $Y(1|2)$ are eigenvectors of the zero-modes $T^{(0)}_{jj}$, $j=1,2,3$:
\bea\label{cartan}
&&T^{(0)}_{11}\,\Phi_{a,b}(\bar u,\bar v) = (\lambda_1^{(0)}-(-1)^{[1]}a)\,\Phi_{a,b}(\bar u,\bar v)\,,\quad\\
&&T^{(0)}_{22}\,\Phi_{a,b}(\bar u,\bar v) = (\lambda_2^{(0)}+(-1)^{[2]}(a-b))\,\Phi_{a,b}(\bar u,\bar v)\,,\quad\\
&&T^{(0)}_{33}\,\Phi_{a,b}(\bar u,\bar v) = (\lambda_3^{(0)}+(-1)^{[3]}b)\,\Phi_{a,b}(\bar u,\bar v)\,.\quad
\eea
Moreover, if the Bethe equations
\bea
\frac{\lambda_2(u_j)}{\lambda_1(u_j)} &=& \frac{f_{[1]}(\bar u_j,u_j)}{f_{[2]}(u_j,\bar u_j)}\ \frac1{f_{[2]}(\bar v,u_j)}\,,\quad j=1,2,...,a\,, \\
\frac{\lambda_3(v_j)}{\lambda_2(v_j)} &=& f_{[2]}(v_j,\bar u)\ \frac{f_{[2]}(\bar v_j,v_j)}{f_{[3]}(v_j,\bar v_j)}\,,\quad j=1,2,...,b\,,
\eea
are obeyed, the BVs $\Phi_{a,b}(\bar u,\bar v)$ are eigenvectors of the transfer matrix \eqref{transfer.mat}:
\bea\label{on.shell}
t(z)\,\Phi_{ab}(\bar u,\bar v) &=& \tau(z|\bar u,\bar v)\,\Phi_{ab}(\bar u,\bar v)\,,
\\
\tau(z|\bar u,\bar v) &=& (-1)^{[1]}\,\lambda_1(z)\,f_{[1]}(\bar u,z)+(-1)^{[2]}\,\lambda_2(z)\,f_{[2]}(z,\bar u)
f_{[2]}(\bar v,z)
\nonu
&&+(-1)^{[3]}\,\lambda_3(z)\,f_{[3]}(z,\bar v).
\eea
The functions $f_{[j]}(u,v)$ are defined as
\be{def-f}
f_0(u,v)=1+g(u,v)=f(u,v)\quad \mb{and}\quad f_1(u,v)=1-g(u,v)=f(v,u)\,.
\ee
\end{prop}

 Note that due to the grading, in the $Y(2|1)$ case   (resp. $Y(1|2)$ case), the ratio of $f$ functions in the r.h.s. of the  
second (resp. first) Bethe equation cancels, and we are left with a free fermion equation.

Using the super-trace formula, one can show the following symmetry property
\begin{prop}\label{prop.sym}
Let $\sigma_j=\sigma_{j,j+1}$ be the transposition between $j$ and $j+1$. Then, for the BVs of $Y(2|1)$, we have
\be{sym}
 \Phi_{a,b}(\bar u,\bar v)  = \Phi_{a,b}(\bar u^{\sigma_j},\bar v) \mb{and}
 \Phi_{a,b}(\bar u,\bar v)  = \Phi_{a,b}(\bar u,\bar v^{\sigma_j})\,.
\ee

In the case of $Y(m|n)$, denoting $\bar t^{(1)},...,\bar t^{(m+n-1)}$ the sets of Bethe parameters,
relation \eqref{sym} will apply to the any set $\bar t^{(j)}$  $j=1,2,...m+n-1$, provided one normalizes the super-trace with $H(\bar t^{(m)})$ .
\end{prop}
\prof We prove the property for $\sigma_{12}$ and $Y(2|1)$, but it extends trivially to $\sigma_j$ and $Y(m|n)$.
We first write $T_1(u_1)T_2(u_2) = R_{12}^{-1}\,T_2(u_2)T_1(u_1)R_{12}$. Then, iterative use of the Yang--Baxter equation shows that
\be{}
R_{12}\,\RR_{123\dots a+b} = \RR_{213\dots a+b}\,R_{12}.
\ee
From cyclicity of the super-trace, one gets
\begin{multline}\label{C-st}
\Phi_{a,b}(\bar u,\bar v) = (-1)^{b(b+1)/2}\, {\str}_{1\cdots a+b} \Big[ T_2(u_2)T_1(u_1)T_3(u_3)\cdots T_a(u_a)T_{a+1}(v_1)\cdots T_{a+b}(v_b)\,
\RR_{213\dots a+b}\\
\times \EE_{32}^{(a+b)}\cdots \EE_{32}^{(a+1)}\EE_{21}^{(a)}\cdots \EE_{21}^{(3)}\,R_{12}\,\EE_{21}^{(2)}\EE_{21}^{(1)}\,R_{12}^{-1}\Big]\Omega.
\end{multline}
Finally a direct calculation, using the explicit form of the $R$-matrix, shows that
\be{}
R_{12}\,\EE_{21}^{(2)}\,\EE_{21}^{(1)}\,R_{12}^{-1} = \EE_{21}^{(1)}\,\EE_{21}^{(2)}.
\ee
Then, after relabeling of the auxiliary spaces 1 and 2, one recognizes in the right-hand-side $\Phi_{a,b}(\bar u^{t_1},\bar v) $, which proves the first relation of proposition \ref{prop.sym}.

The proof of the second relation follows the same lines, the only difference lies in the grade of $\EE_{32}$, which leads to
\be{}
R_{a+1,a+2}\,\EE_{32}^{(a+2)}\,\EE_{32}^{(a+1)}\,R_{a+1,a+2}^{-1}
=  -\frac{f(v_{2},v_1)}{f(v_1,v_{2})}\,\EE_{32}^{(a+1)}\,\EE_{32}^{(a+2)}=  \frac{h(v_{2},v_1)}{h(v_1,v_{2})}\,\EE_{32}^{(a+1)}\,\EE_{32}^{(a+2)}.
\ee
This coefficient cancels the one coming from the normalization factor $H(\bar v)$.\qed

\subsection{Comparison with Tarasov--Varchenko trace formula}

Starting from \cite{TV}, and generalizing to the superalgebra case, one would get Bethe vectors defined as\footnote{However, be careful that the normalisation of the $R$-matrix is different in this paper.}
\be{BV-TV}
\cW_{ab}(\bar u,\bar v) = {\str} \Big( T_1(u_1)...T_a(u_a)T_{a+1}(v_1)...T_{a+b}(v_b)
\,\cR_{1...a+b}\ (\EE_{21})^{\otimes a}\ (\EE_{32})^{\otimes b}\Big)
\ee
with
\bea
\cR_{1...a+b}&=&\prod_{1\leq i<j\leq a+b}^{\to} R_{ji}
=\left(\prod_{a+1\leq i<j\leq a+b}^{\to} R_{ji}\right)\wt\RR_{1...a+b}\left(\prod_{1\leq i<j\leq a}^{\to} R_{ji}\right)
\\
\wt\RR_{1...a+b} &=& (R_{a+b,a}\dots R_{a+b,1})\cdots(R_{a+2,a}\dots R_{a+2,1})(R_{a+1,a}\dots R_{a+1,1})
\\
&\equiv&\RR_{1...a,a+b...a+1}
\eea
where $\RR_{1...a+b}$ is defined in \eqref{def:RR}.
The form of $\cW_{ab}(\bar u,\bar v)$ is different from the one of $\Phi_{ab}(\bar u,\bar v)$, given in \cite{BR1}, and reproduced in \eqref{def:Phi-str1}. However, we have the following.
\begin{prop}
In $Y(2|1)$, we have
\be{W-phi}
\cW_{ab}(\bar u,\bar v) =  \prod_{1\leq i<j\leq a}f(u_j,u_i)\ \prod_{1\leq i<j\leq b}g(v_j,v_i)\
\Phi_{ab}(\bar u^*,\bar v^*)
\ee
where for any set $\bar w=\{w_1,w_2,...,w_b\}$, we introduced the conjugate one $\bar w^*=\{w_b,...,w_2,w_1\}$.

Relation \eqref{W-phi} trivially generalizes to the $Y(m|n)$ case.
\end{prop}
\prof A direct calculation shows that
$$
\left(\prod_{1\leq i<j\leq a}^{\to} R_{ji}\right)\,(\EE_{21})^{\otimes a}
= \prod_{1\leq i<j\leq a}f(u_j,u_i)\,(\EE_{21})^{\otimes a}.
$$
Now, from the relation
$$
T_{a+1}(v_1)...T_{a+b}(v_b)\left(\prod_{a+1\leq i<j\leq a+b}^{\to} R_{ji}\right)
=\left(\prod_{a+1\leq i<j\leq a+b}^{\to} R_{ji}\right)T_{a+b}(v_b)...T_{a+1}(v_1),
$$
the cyclicity property of the super-trace and the property
$$
(\EE_{32})^{\otimes b}\left(\prod_{a+1\leq i<j\leq a+b}^{\to} R_{ji}\right)
=\prod_{1\leq i<j\leq b}f(v_j,v_i)\ (\EE_{32})^{\otimes b}
$$
we get
\bea
\cW_{ab}(\bar u,\bar v) &=& \prod_{1\leq i<j\leq a}f(u_j,u_i)\ \prod_{1\leq i<j\leq b}f(v_j,v_i)
\nonu
&&\times\ {\str} \Big( T_1(u_1)...T_a(u_a)T_{a+b}(v_b)...T_{a+1}(v_1)
\,\wt\RR_{1...a+b}\ (\EE_{21})^{\otimes a}\ (\EE_{32})^{\otimes b}\Big).
\eea
Finally, one relabels the spaces $a+1,...,a+b$ into $a+b,...,a+1$ and the Bethe parameters $v_j$ accordingly. One recovers the formula \eqref{W-phi} after division by $H(\bar v)$.\qed

\section{First recursion formula for $Y(2|1)$ and $Y(1|2)$ BVs\label{sect:1strecur}}
\subsection{$Y(2|1)$ case}
We focus now on the case of $Y(2|1)$.  Recall that the grading is given by $[1]=[2]=0$ and $[3]=1$.

\begin{prop}\label{prop:recur12}
The $Y(2|1)$ Bethe vectors obey the following recursion relation:
\bea
\Phi_{ab}(\bar u,\bar v) &=& T_{12}(u_1)\, \Phi_{a-1,b}(\bar u_1,\bar v)
\nonu
&&+\sum_{j=1}^b \lambda_2(v_j)\,
g(v_j,u_1)\,f(v_j,\bar u_1)
g(\bar v_j, v_j)\,
 T_{13}(u_1)\,\Phi_{a-1,b-1}(\bar u_1,\bar v_j),
\label{eq:phi-T12}
\eea
and we recall that $\bu_1=\bu\setminus \{u_1\}$.
\end{prop}
\prof One extracts the space 1 dependence from the super-trace formula. To do it, we need the following formulas:
\be{}
R_{21}(v,u)\,\EE_{32}^{(2)}\,\EE_{21}^{(1)} = \EE_{32}^{(2)}\,\EE_{21}^{(1)} +g(v,u)\EE_{31}^{(1)}\,\EE_{22}^{(2)}
\mb{and}
R_{21}(v,u)\,\EE_{32}^{(2)}\,\EE_{31}^{(1)} = f(u,v)\EE_{32}^{(2)}\,\EE_{31}^{(1)} .
\ee
Applying these relations recursively on the matrices $R_{a+j,1}$, $j=1,2,...,b$, we get
\begin{align}
H(\bar v)\,\Phi_{ab}(\bar u,\bar v)&=T_{12}(u_1)\, \Phi_{a-1,b}(\bar u_1,\bar v)
-\,T_{13}(u_1)\,\sum_{j=1}^b (-1)^{b(b-1)/2+j}
g(v_j,u_1)\prod_{1\leq p<j}f(u_1,v_p)\qquad \label{recur-part1}\\
&\times
{\str}_{2\cdots a+b} \left[ T_2(u_2)\cdots T_{a+b}(v_b)\,
\RR_{2\dots a+b}\, \EE_{32}^{(a+b)}\cdots \EE_{22}^{(a+j)}\cdots \EE_{32}^{(a+1)}\EE_{21}^{(a)}\cdots \EE_{21}^{(2)}\right]\Omega
\nonumber
\end{align}
where we have used
$\RR_{1\dots a+b}=\RR_{2\dots a+b}\,R_{a+1,1}\dots R_{a+b,1}$.

It remains to eliminate the generator $\EE_{22}^{(a+j)}$, in the second term of eq. \eqref{recur-part1}, which is done by performing the super-trace on the space $a+j$. For such a purpose, one first notes that
\be{}
R_{a+j,k}(v,u)\,\EE_{22}^{(a+j)}\,\EE_{21}^{(k)} = (1 +g(v,u))\,\EE_{22}^{(a+j)}\,\EE_{21}^{(k)}.
\ee
This shows that the elimination of the generator $\EE_{22}^{(a+j)}$, will produce a generator $T_{22}$. To get rid of this generator, one needs to move it to the right towards $\Omega$, to get a $\lambda_2$ function. Moving $T_{22}(v_j)$ to the right is done thanks to the commutation relations\footnote{Because of the super-trace in space $p>a+j$ with the generator $\EE_{32}$, one knows by cyclicity that in $T_p(v)$ only the generators  $T_{2k}(v)$ will matter.}
\be{t22}
T_{22}(x)\,T_{2k}(y) = g(x,y)\,T_{2k}(x)T_{22}(y)+f(y,x)\,T_{2k}(y)T_{22}(x).
\ee
Altogether, this implies that eliminating the generator $\EE_{22}^{(a+j)}$ (and summation on $j=1,...,b$), we will get a sum
$\sum_{\ell=1}^b \lambda_2(v_\ell) (...)$. To compute precisely the form of these terms, we will use the symmetry property \ref{prop.sym}, see below.

First, we compute the term corresponding to $\lambda_2(v_1)$. Because of the relations \eqref{t22} and the order in the product of $T$'s in \eqref{recur-part1}, it is clear that $\lambda_2(v_1)$ can be obtained only through $T_{22}(v_1)$, which in turn means that we need to determine how to get $\EE_{22}^{(a+1)}$ after $\EE_{22}^{(a+j)}$, $j=1,...,b$, goes through $\RR_{2\dots a+b}$.

We fix $j$, and look at $\EE_{22}^{(a+j)}$. From the relations
\bea
R_{a+p,q}\,\EE_{32}^{(a+p)}\,\EE_{21}^{(q)} &=& \EE_{32}^{(a+p)}\,\EE_{21}^{(q)} +g(v_p,u_q)\,\EE_{22}^{(a+p)}\,\EE_{31}^{(q)}\,,
\quad 1\leq q \leq a\ ;\ j<p\leq b \\
R_{a+j,k}\,\EE_{22}^{(a+j)}\,\EE_{21}^{(k)} &=& 
f(v_j,u_k)\,\EE_{22}^{(a+j)}\,\EE_{21}^{(k)}\,,
\quad 1\leq q \leq a  \\
R_{a+p,q}\,\EE_{22}^{(a+p)}\,\EE_{31}^{(q)} &=& \EE_{22}^{(a+p)}\,\EE_{31}^{(q)} +g(v_p,u_q)\,\EE_{32}^{(a+p)}\,\EE_{21}^{(q)} \,,
\quad 1\leq q \leq a\ ;\ j<p\leq b
\eea
it is clear that $\EE_{22}^{(a+j)}$ produces terms $\EE_{22}^{(a+p)}$ with $p\geq j$ only.
It implies that there is only one way to get
$\EE_{22}^{(a+1)}$ after going through $\RR_{2\dots a+b}$, and we get
\be{}
f(v_1,\bar u_1)\,{\str}_{2\cdots a+b} T_2(u_2)\cdots T_{a+b}(v_b)\,\EE_{22}^{(a+1)}\,
\RR_{2\dots a, a+2\dots a+b}\, \EE_{32}^{(a+b)}\cdots  \EE_{32}^{(a+2)}\EE_{21}^{(a)}\cdots \EE_{21}^{(2)}\,\Omega\,.
\ee
It remains to perform the super-trace in space $a+1$ and move $T_{22}(v_1)$ towards $\Omega$.
We obtain
\bea
\Phi_{ab}(\bar u,\bar v) &=& T_{12}(u_1)\, \Phi_{a-1,b}(\bar u_1,\bar v)
\nonu
&&+\,T_{13}(u_1)\, \lambda_2(v_1)
g(v_1,u_1) f(v_1,\bar u_1) f(\bar v_1, v_1)\,\frac{H(\bar v_1)}{H(\bar v)}\,\Phi_{a-1,b-1}(\bar u_1,\bar v_1)\nonu
&&+ \sum_{\ell=2}^b  \lambda_2(v_\ell) (...), \label{recur-part2}
\eea
where the dots encode all the terms that contribute to obtain $T_{22}(v_\ell)$ on the right.

In the same way, for $\bar v^\sigma =\{v_j,\bar v_j\}$, we have
\bea
\Phi_{ab}(\bar u,\bar v^\sigma) &=& T_{12}(u_1)\, \Phi_{a-1,b}(\bar u_1,\bar v^\sigma)
\nonu
&&+\,T_{13}(u_1)\, \lambda_2(v_j)
g(v_j,u_1) f(v_j,\bar u_1) f(\bar v_j, v_j)\,\frac{H(\bar v_j)}{H(\bar v^\sigma)}\,\Phi_{a-1,b-1}(\bar u_1,\bar v_j)\nonu
&&+ \sum_{\ell\neq j}  \lambda_2(v_\ell) (...) . \label{recur-part3}
\eea

But $\bar v^\sigma$ is deduced from $\bar v$ from action of the permutation  $\sigma=\sigma_{12}\, \sigma_{23}\dots \sigma_{j-2,j-1}\, \sigma_{j-1,j}$
so that from property \ref{prop.sym}, $\Phi_{ab}(\bar u,\bar v^\sigma) =\Phi_{ab}(\bar u,\bar v)$. It remains to compute
$$
\frac{H(\bar v_j)}{H(\bar v^\sigma)} =h(\bar v_j,v_j)
$$
to get the coefficient of $\lambda_2(v_j)$ in the recursion relation.\qed

\subsection{$Y(1|2)$ case}
We focus now on the case of $Y(1|2)$. The grading is given by $\wt{[1]}=0$ and $\wt{[2]}=\wt{[3]}=1$.
The super-trace formula read
\bea
\wt\Phi_{a,b}(\bar u,\bar v) &=&  \frac{(-1)^{a}}{H(\bar u)}\,
{\str}_{1\cdots a+b} \Big[\wt T_1(u_1)\cdots\wt T_a(u_a)\wt T_{a+1}(v_1)\cdots\wt T_{a+b}(v_b)\,\RR_{1\dots a+b}
\nonu
&&\qquad\qquad\qquad\qquad\qquad\times\,
\, \EE_{21}^{(1)}\cdots \EE_{21}^{(a)}\EE_{32}^{(a+1)}\cdots \EE_{32}^{(a+b)}\Big]\Omega\,,
\label{def:Phitilde-str}
\eea
where now $\EE_{21}$ is odd while $\EE_{32}$ is even.
We have put a tilde on BVs to distinguish the $Y(1|2)$ BVs from the $Y(2|1)$ ones.

\begin{prop}\label{prop:recur12-gl21}
The $Y(1|2)$ Bethe vectors obey the following recursion relation:
\bea
\wt\Phi_{ab}(\bar u,\bar v) &=&h(\bar u_1,u_1)^{-1}\,\wt T_{12}(u_1)\, \wt\Phi_{a-1,b}(\bar u_1,\bar v)
\label{eq:phitilde-T12}\\
& -&\!h(\bar u_1,u_1)^{-1}\,\sum_{j=1}^b \wt\lambda_2(v_j)\,
g(u_1,v_j)\,f(\bar u_1,v_j)\, f(\bar v_j,v_j)\,\wt T_{13}(u_1)\,\wt\Phi_{a-1,b-1}(\bar u_1,\bar v_j).
\nonumber
\eea
\end{prop}
\prof
The proof goes along the same line as for proposition \ref{prop:recur12},
taking into account the difference between the gradings and the normalisation factor, which now depends on $\bar u$.
For instance, the exchange relation \eqref{t22} now reads
\be{t22-gl12}
\wt T_{22}(x)\,\wt T_{2k}(y) = g(y,x)\,\wt T_{2k}(x)\wt T_{22}(y)+f(x,y)\,\wt T_{2k}(y)\wt T_{22}(x).
\ee
\qed

\section{Second recursion formula for $Y(2|1)$ and $Y(1|2)$ BVs\label{sect:2ndrecur}}
To get the second recursion formula, one needs to use the morphism $\vph$ defined in section \ref{sect:morphism}.
To clarify the presentation, we use the following notation:
 
\be{vph-tilde}
\begin{array}{ccccc} &Y(2|1)\ \to\ Y(1|2)
&&&Y(1|2)\ \to\ Y(2|1)\\[1ex]
\vph:\ &\begin{cases}
T_{12}(u) \ \to\ -\wt T_{23}(u)\\
T_{23}(u) \ \to\ \wt T_{12}(u)\\
T_{13}(u) \ \to\ \wt T_{13}(u)\\
\lambda_{2}(u) \ \to\ -\wt \lambda_{2}(u)
\end{cases}
&\mbox{and}&\wt\vph: &\quad\begin{cases}
\wt T_{12}(u) \ \to\  T_{23}(u)\\
\wt T_{23}(u) \ \to\  -T_{12}(u)\\
\wt T_{13}(u) \ \to\  T_{13}(u)
\\
\wt\lambda_{2}(u) \ \to\  -\lambda_{2}(u)\end{cases}
\end{array}
\ee
where we have explicitly written the image of the generators needed in the following.

\begin{lemma}\label{lem:vph}
The isomorphisms $\vph$ and $\wt\vph$ provide the following relations between $Y(1|2)$ BVs and $Y(2|1)$ ones:
 
\be{phi-BV}
\vph\Big(\Phi_{ab}(\bar u,\bar v)\Big) = \,\wt\Phi_{ba}(\bar v,\bar u)
\mb{and}
\wt\vph\Big(\wt\Phi_{ab}(\bar u,\bar v)\Big) = \,\Phi_{ba}(\bar v,\bar u).
\ee
\end{lemma}
\prof
We start with an on-shell BV of $Y(1|2)$ and apply $\wt\vph$ to the equality \eqref{on.shell}, written in $Y(1|2)$. 
 Since $\wt\vph$ is compatible with the supertrace, one has $\wt\vph(\wt t(z))=t(z)$. This shows that
$\wt\vph\Big(\wt\Phi_{ab}(\bar u,\bar v)\Big)$ is an eigenvector of $t(z)$ in $Y(2|1)$. Acting with $\wt\vph$ on the relations \eqref{cartan}, and using \eqref{vph-tilde}, it shows that it is proportional to $\Phi_{ba}(\bar v,\bar u)$.
In the same way, we show that $\vph\Big(\Phi_{ab}(\bar u,\bar v)\Big)$  is proportional to $\wt\Phi_{ba}(\bar v,\bar u)$.

To fix the normalisation in equality \eqref{phi-BV}, we consider a specific coefficient  in the super-trace formula. To get this coefficient, we will use
\bea
&\mbox{In }Y(2|1):& {\str}(T(u)\,\EE_{21})=T_{12}(u)\quad;\quad {\str}(T(u)\,\EE_{32})=-T_{23}(u)\\
&\mbox{In }Y(1|2):& {\str}(\wt T(u)\,\EE_{21})=-\wt T_{12}(u)\quad;\quad {\str}(\wt T(u)\,\EE_{32})=-\wt T_{23}(u).
\eea
Considering \eqref{def:Phi-str2} written in $Y(2|1)$, the coefficient of $T_{12}(\bar u)\, T_{23}(\bar v)$ in $\Phi_{ab}(\bar u,\bar v)$ is
$H(\bar v)^{-1}$.
Through the action of $\vph$, it provides a term  $ (-1)^a\,\wt T_{23}(\bar u)\, \wt T_{12}(\bar v)$. To get the corresponding term in
$\wt\Phi_{ab}(\bar u,\bar v)$, we use the relation \eqref{rtt} to rewrite \eqref{eq:phitilde-T12} as
\bea
\wt\Phi_{a,b}(\bar u,\bar v) &=& \frac{(-1)^{a}}{H(\bar u)}\, {\str}_{1\cdots a+b} \Big[\RR_{1\dots a+b}\, \wt T_{a+1}(v_1)\cdots \wt T_{a+b}(v_b)\wt T_1(u_1)\cdots \wt T_a(u_a)
\nonu
&&\qquad\qquad\qquad\qquad\qquad\times\,
\, \EE_{21}^{(1)}\cdots \EE_{21}^{(a)}\EE_{32}^{(a+1)}\cdots \EE_{32}^{(a+b)}\Big]\Omega\,.
\label{def:Phi-str3}
\eea
Then the coefficient of $\wt T_{23}(\bar u)\, \wt T_{12}(\bar v)$ in $\wt\Phi_{ba}(\bar v,\bar u)$ is $(-1)^aH(\bar v)^{-1}$, leading
 to the first relation in \eqref{phi-BV}.

Finally, the coefficient of $\wt T_{12}(\bar u)\, \wt T_{23}(\bar v)$ in $\wt\Phi_{ab}(\bar u,\bar v)$ is $(-1)^bH(\bar u)^{-1}$, which, using $\wt\vph$, is sent to
$ H(\bar u)^{-1}\, T_{23}(\bar u)\, T_{12}(\bar v)$. The comparison with the coefficient of $T_{23}(\bar u)\, T_{12}(\bar v)$ in
$\Phi_{ba}(\bar v,\bar u) $, which is $H(\bar u)^{-1}$, obtained from \eqref{def:Phi-str3}, provides the second relation of \eqref{phi-BV}.
\qed

%
\begin{prop}\label{prop:recur23}
The $Y(2|1)$ Bethe vectors obey the following recursion relation:
\bea
\Phi_{ab}(\bar u,\bar v) &=& h(\bar v_1,v_1)^{-1}\,T_{23}(v_1)\, \Phi_{a,b-1}(\bar u,\bar v_1)
\label{eq:recur23} \\
&+&h(\bar v_1,v_1)^{-1}\,\sum_{j=1}^a \lambda_2(u_j)\,
g(v_1,u_j)\,f(\bar v_1,u_j)\, f(\bar u_j, u_j)\, T_{13}(v_1)\,\Phi_{a-1,b-1}(\bar u_j,\bar v_1).
\nonumber
\eea
The $Y(1|2)$ Bethe vectors obey the following recursion relation:
\bea
\wt\Phi_{ab}(\bar u,\bar v) &=& -\wt T_{23}(v_1)\, \wt\Phi_{a,b-1}(\bar u,\bar v_1)
\nonu
&&\label{eq:recur23tilde}
 -\sum_{j=1}^a \wt\lambda_2(u_j)\,g(u_j,v_1)\,f(u_j,\bar v_1)\, g(\bar u_j,u_j)
\, \wt T_{13}(v_1)\,\wt\Phi_{a-1,b-1}(\bar u_j,\bar v_1).
\nonumber
\eea
\end{prop}
\prof One starts with the recursion formula for $Y(1|2)$ BVs, as given in
proposition \ref{prop:recur12-gl21} and rewritten as
\bea
\wt\Phi_{ba}(\bar v,\bar u) &=& h(\bar v_1,v_1)^{-1}\,\wt T_{12}(v_1)\, \wt\Phi_{b-1,a}(\bar v_1,\bar u)
 \\
&-&h(\bar v_1,v_1)^{-1}\,\sum_{j=1}^a \wt \lambda_2(u_j)\,
g(v_1,u_j)\,f(\bar v_1,u_j)\, f(\bar u_j,u_j)\, \wt T_{13}(v_1)\,\wt\Phi_{a-1,b-1}(\bar v_1,\bar u_j)\,.
\nonumber
\eea
Now, applying the isomorphism $\vph$, we get \eqref{eq:recur23}.

In the same way, starting from \eqref{eq:phi-T12} written for $\Phi_{ba}(\bar v,\bar u)$ and applying $\wt\vph$, we get \eqref{eq:recur23tilde}.
\qed

Note that applying again the morphism $\vph$ (resp. $\wt\vph$) on relation \eqref{eq:recur23} (resp. \eqref{eq:recur23tilde}), one gets back to the recursion relation \eqref{eq:phitilde-T12}  (resp. \eqref{eq:phi-T12}).

\section{Dual Bethe vectors\label{sect:dual}}
Using the antimorphism $\psi$, one can map BVs into dual BVs, that are left-eigenvectors of the transfer matrix (when on-shell).
Again, to lighten the presentation, we focus on $Y(2|1)$ and $Y(1|2)$ BVs, but the technique applies also to the  $Y(m|n)$ case.
We define the dual Bethe vectors as
\be{dualBVs}
\Psi_{ab}(\bar u,\bar v) = \psi\big(\Phi_{ab}(\bar u^*,\bar v^*)\big).
\ee
Recall that for any set $\bar w =\{w_1,w_2,...,w_a\}$, we define its conjugate set as $\bar w^*=\{w_a,...,w_2,w_1\}$.
\subsection{Supertrace formula}
\begin{prop}
The $Y(2|1)$  dual BVs admit the following super-trace expression:
\bea
\Psi_{ab}(\bar u,\bar v)&=& \frac{(-1)^{b(b-1)/2}}{H(\bar v^*)}\,
\Omega^\dag\,{\str}_{1\cdots a+b} \Big[ T_1(u_1)\cdots T_a(u_a)T_{a+1}(v_1)\cdots T_{a+b}(v_b)\,\RR_{1\dots a+b}
\nonu
&&\qquad\qquad\qquad\qquad\qquad\times\,
\, \EE_{12}^{(1)}\cdots \EE_{12}^{(a)}\EE_{23}^{(a+1)}\cdots \EE_{23}^{(a+b)}\Big]
\label{str-dual21}
\eea
where we noted $\psi(\Omega)=\Omega^\dag$ and
\be{H*}
H(\bar v^*)=\prod_{1\leq j<k\leq b}h(v_j,v_k) .
\ee

In the same way, $Y(1|2)$  dual BVs admit the following super-trace expression:
\bea
\wt\Psi_{ab}(\bar u,\bar v)&=& \frac{(-1)^{a(a-1)/2}}{H(\bar u^*)}\, \wt\Omega^\dag\,{\str}_{1\cdots a+b} \Big[\wt T_1(u_1)\cdots\wt T_a(u_a)\wt T_{a+1}(v_1)\cdots\wt T_{a+b}(v_b)\,\RR_{1\dots a+b}
\nonu
&&\qquad\qquad\qquad\qquad\qquad\times\,
\, \EE_{12}^{(1)}\cdots \EE_{12}^{(a)}\EE_{23}^{(a+1)}\cdots \EE_{23}^{(a+b)}\Big]
\label{str-dual12}
\eea
with $\wt\psi(\wt\Omega)=\wt\Omega^\dag$.
\end{prop}
\prof
Applying $\psi$ to the expression \eqref{def:Phi-str2}, one gets
\bea
\Psi_{ab}(\bar u^*,\bar v^*) &=& \frac{(-1)^{b}}{H(\bar v)}\, \Omega^\dag\,{\str}_{1\cdots a+b} \Big[ \psi\big(T_{a+b}(v_b)\big)\cdots
\psi\big(T_{a+1}(v_1)\big)\psi\big(T_a(u_a)\big)\cdots \psi\big(T_1(u_1)\big)\,\RR_{1\dots a+b}
\nonu
&&\qquad\qquad\qquad\qquad\qquad\times\,
\, \EE_{21}^{(1)}\cdots \EE_{21}^{(a)}\EE_{32}^{(a+1)}\cdots \EE_{32}^{(a+b)}\Big]
\\
 &=& \frac{(-1)^{b}}{H(\bar v)}\, \Omega^\dag\,{\str}_{1\cdots a+b} \Big[ T_{a+b}(v_b)^t\cdots
T_{a+1}(v_1)^t\,T_a(u_a)^t\cdots T_1(u_1)^t\,\RR_{1\dots a+b}
\nonu
&&\qquad\qquad\qquad\qquad\qquad\times\,
\, \EE_{21}^{(1)}\cdots \EE_{21}^{(a)}\EE_{32}^{(a+1)}\cdots \EE_{32}^{(a+b)}\Big]
\\
 &=& \frac{(-1)^{b}}{H(\bar v)}\, \Omega^\dag\,{\str}_{1\cdots a+b} \Big[ \Big(T_{a+b}(v_b)\cdots
T_{a+1}(v_1)\,T_a(u_a)\cdots T_1(u_1)\Big)^{t_1\cdots t_{a+b}}\,\RR_{1\dots a+b}
\nonu
&&\qquad\qquad\qquad\qquad\qquad\times\,
\, \EE_{21}^{(1)}\cdots \EE_{21}^{(a)}\EE_{32}^{(a+1)}\cdots \EE_{32}^{(a+b)}\Big]
\\
 &=& \frac{(-1)^{b}}{H(\bar v)}\, \Omega^\dag\,{\str}_{1\cdots a+b} \Big[T_{a+b}(v_b)\cdots
T_{a+1}(v_1)\,T_a(u_a)\cdots T_1(u_1)\,
\nonu
&&\qquad\qquad\qquad\qquad\qquad\times\,
\Big(\RR_{1\dots a+b}\, \EE_{21}^{(1)}\cdots \EE_{21}^{(a)}\EE_{32}^{(a+1)}\cdots \EE_{32}^{(a+b)}\Big)^{t^3_1\cdots t^3_{a+b}}\Big],
\eea
where to get the last step we have used relation \eqref{transpo-trace} for the cube of the transposition $(\cdot)^{t^3}$ and the fact that
$(\cdot)^{t^4}=id$. Then, using $\EE_{21}^{t^3}=\EE_{12}$ and $\EE_{32}^{t^3}=-\EE_{23}$, we get
\bea
{H(\bar v)}\,\Psi_{ab}(\bar u^*,\bar v^*) &=& \Omega^\dag\,{\str}_{1\cdots a+b} \Big[T_{a+b}(v_b)\cdots
T_{a+1}(v_1)\,T_a(u_a)\cdots T_1(u_1)\,
\nonu
&&\qquad\qquad\qquad\qquad\times\,\EE_{12}^{(1)}\cdots \EE_{12}^{(a)}\EE_{23}^{(a+1)}\cdots \EE_{23}^{(a+b)}
\, \RR_{1\dots a+b}^{t^3_1\cdots t^3_{a+b}}\Big].
\eea
Now, relabeling the spaces $(1,2,...,a)$ to $(a,...,2,1)$ and $(a+1,a+2,...,a+b)$ to $(a+b,...,a+2,a+1)$, and the Bethe parameters accordingly, we get
\bea
H(\bar v^*)\,\Psi_{ab}(\bar u,\bar v) &=&  \Omega^\dag\,{\str}_{1\cdots a+b} \Big[T_{a+1}(v_1)\cdots
T_{a+b}(v_b)\,T_1(u_1)\cdots T_a(u_a)\,
\nonu
&&\qquad\qquad\times\,\EE_{12}^{(a)}\cdots \EE_{12}^{(1)}\EE_{23}^{(a+b)}\cdots \EE_{23}^{(a+1)}
\, \RR_{a\dots1, a+b\dots a+1}^{t^3_1\cdots t^3_{a+b}}\Big]
\\
&=& \Omega^\dag\,{\str}_{1\cdots a+b} \Big[\RR_{1\dots a+b}T_{a+1}(v_1)\cdots
T_{a+b}(v_b)\,T_1(u_1)\cdots T_a(u_a)\,
\nonu
&&\qquad\qquad\times\,\EE_{12}^{(a)}\cdots \EE_{12}^{(1)}\EE_{23}^{(a+b)}\cdots \EE_{23}^{(a+1)}
\, \Big],
\eea
where in the last step we have used the property $ \RR_{a\dots1, a+b\dots a+1}^{t^3_1\cdots t^3_{a+b}}=\RR_{1\dots a+b}$ and cyclicity of the super-trace. Notice that $H(\bar v)$ and $H(\bar v^*)$ are not equal: compare the definition \eqref{def:H} of $H(\bar v)$ with
\eqref{H*}.

Finally, using repetitively relation \eqref{rtt}, one proves that
$$
\RR_{1\dots a+b}T_{a+1}(v_1)\cdots T_{a+b}(v_b)\,T_1(u_1)\cdots T_a(u_a) =
T_1(u_1)\cdots T_a(u_a)T_{a+1}(v_1)\cdots T_{a+b}(v_b)\,\RR_{1\dots a+b}
$$
and one gets \eqref{str-dual21} after reordering the tensor product $\EE_{12}^{(a)}\cdots \EE_{12}^{(1)}\EE_{23}^{(a+b)}\cdots \EE_{23}^{(a+1)}$.
The same calculation starting from \eqref{def:Phitilde-str}, with now $\EE_{21}^{t^3}=-\EE_{12}$ and $\EE_{32}^{t^3}=\EE_{23}$, leads to
\eqref{str-dual12}.
\qed

\subsection{Recursion formulas}
\begin{prop}
The $Y(2|1)$ dual BVs obey the following recursion relations
\bea
\Psi_{ab}(\bar u,\bar v) &=& \Psi_{a-1,b}(\bar u_a,\bar v)\,T_{21}(u_a)
\nonu
&+&(-1)^{b-1}\,\sum_{j=1}^b \lambda_2(v_j)\,
g(v_j,u_a)\,f(v_j,\bar u_a)\,g(\bar v_j,v_j)
\, \Psi_{a-1,b-1}(\bar u_a,\bar v_j)\,T_{31}(u_a),
\label{21dual-recur21}
\\
\Psi_{ab}(\bar u,\bar v) &=& (-1)^{b-1}\,h(\bar v_b,v_b)^{-1}\, \Psi_{a,b-1}(\bar u,\bar v_b)\,T_{32}(v_b)\,
\label{21dual-recur32} \\
&+&(-1)^{b-1}\,h(\bar v_b,v_b)^{-1}\, \sum_{j=1}^a \lambda_2(u_j)\,
g(v_b,u_j)\,f(\bar v_b,u_j)\, f(\bar u_j, u_j)\,\Psi_{a-1,b-1}(\bar u_j,\bar v_b)\, T_{31}(v_b).
\nonumber
\eea
The $Y(1|2)$ dual BVs obey the following recursion relations
\bea
\wt\Psi_{ab}(\bar u,\bar v) &=& (-1)^{a-1}\, h(\bar u_a,u_a)^{-1}\,\wt\Psi_{a-1,b}(\bar u_a,\bar v)\,\wt T_{21}(u_a)
\label{12dual-recur21}\\
& -&(-1)^{a-1}\,h(\bar u_a,u_a)^{-1}\, \sum_{j=1}^b \wt\lambda_2(v_j)\,
g(u_a,v_j)\,f(\bar u_a,v_j)\, f(\bar v_j, v_j)\,\wt\Psi_{a-1,b-1}(\bar u_a,\bar v_j)\,\wt T_{31}(u_a),
\nonu
\wt\Psi_{ab}(\bar u,\bar v) &=& - \wt\Psi_{a,b-1}(\bar u,\bar v_b)\,\wt T_{32}(v_b)\,
\label{12dual-recur32} \\
& -&(-1)^{a-1}\,\sum_{j=1}^a \wt\lambda_2(u_j)\,
g(u_j,v_b)\,f(u_j,\bar v_b)
g(\bar u_j,u_j)
\ \wt\Psi_{a-1,b-1}(\bar u_j,\bar v_b)\,\wt T_{31}(v_b).
\nonumber
\eea
\end{prop}
\prof
Applying $\psi$ to the relation \eqref{eq:phi-T12}, we get
\bea
\Psi_{ab}(\bar u^*,\bar v^*) &=& \Psi_{a-1,b}({\bar u_1}^*,\bar v^*)\,T_{21}(u_1)
\nonu
&+&\sum_{j=1}^b (-1)^{b-1}\,\lambda_2(v_j)\,
g(v_j,u_1)\,f(v_j,\bar u_1)g(\bar v_j,v_j)
 \, \Psi_{a-1,b-1}({\bar u_1}^*,{\bar v_j}^*)\,T_{31}(u_1)
\eea
where ${\bar u_1}^*$ is the conjugate of the set $\bar u_1$ and we remind that $\psi$  acts as
$$
\psi\big(T_{12}(u)\big) =  T_{21}(u)\,,\quad \psi\big(T_{23}(u)\big) = T_{32}(u)\,,\quad \psi\big(T_{13}(u)\big) = T_{31}(u)\,.
$$
 Now, we set $k=b+1-j$,  to get
\bea
\Psi_{ab}(\bar u^*,\bar v^*) &=&  \Psi_{a-1,b}(\bar u^*_1,\bar v^*)\,T_{21}(u_1)
+\sum_{k=1}^b (-1)^{b-1}\,\lambda_2(v_{b+1-k})\,
g(v_{b+1-k},u_1)\,f(v_{b+1-k},\bar u_1)\nonu
&&\qquad\quad \times \,g(\bar v_{b+1-k},v_{b+1-k})
\,\Psi_{a-1,b-1}(\bar u^*_1,\bar v^*_{b+1-k})\,T_{31}(u_1).
\eea
Finally, to get relation \eqref{21dual-recur21}, one relabels the Bethe parameters as $v'_{j}=v_{b+1-j}$ and $u'_{j}=u_{a+1-j}$.

The proof for relation \eqref{21dual-recur32} follows the same lines.
Relations \eqref{12dual-recur21} and \eqref{12dual-recur32} are proven the same way, taking into account the
different $\ZZ_2$-grading.\qed

\subsection{Morphisms on dual vectors}
To be complete, we also provide relations between BVs and dual BVs. As usual, $\Phi$ and $\Psi$ refer to $Y(2|1)$ while
$\wt\Phi$ and $\wt\Psi$ correspond to $Y(1|2)$.
\begin{lemma}
We have the following relations
\bea
&&\psi(\Psi_{ab}(\bar u,\bar v)) = \, \Phi_{ab}(\bar u^*,\bar v^*)\,,\quad
\wt \psi(\wt\Psi_{ab}(\bar u,\bar v)) =\, \wt\Phi_{ab}(\bar u^*,\bar v^*)
\label{psi-CC}\\[1ex]
&&\vph(\Psi_{ab}(\bar u,\bar v)) = \, \wt\Psi_{ba}(\bar v,\bar u)\,,\quad
\wt\vph(\wt\Psi_{ab}(\bar u,\bar v)) = \, \Psi_{ba}(\bar v,\bar u).
\label{phi-CC}
\eea
\end{lemma}
\prof
Relations \eqref{psi-CC} and \eqref{phi-CC} are a direct consequence of lemma \ref{vph-vphtilde}.
For instance
\bea
\vph(\Psi_{ab}(\bar u,\bar v)) = \vph\,\circ\,\psi(\Phi_{ab}(\bar u^*,\bar v^*))
=\wt\psi\,\circ\,\vph(\Phi_{ab}(\bar u^*,\bar v^*))
= \, \wt\psi(\wt\Phi_{ba}(\bar v^*,\bar u^*))
= \, \wt\Psi_{ba}(\bar v,\bar u).
 \eea
The remaining relations are proven the same way.\qed

\section{Explicit expressions for $Y(2|1)$ Bethe vectors\label{sect:explY21}}
From the recursion formulas, one can deduce explicit expressions for BVs and dual BVs. This section is devoted to the proof of the following proposition:
\begin{prop}
In $Y(2|1)$ the Bethe vectors admit the two following explicit expressions
\bea\label{Phi-expl}
\Phi_{a,b}(\bar u,\bar v)&=&
\sum g(\bar v_{\so},\bar u_{\so})\, f(\bar u_{\so},\bar u_{\st})\, g(\bar v_{\st},\bar v_{\so})\, {h(\bar u_{\so},\bar u_{\so})}\;
\TT_{13}(\bar u_{\so})\,T_{12}(\bar u_{\st})\,\TT_{23}(\bar v_{\st})\lambda_2(\bar v_{\so})\,\Omega,
\\
\label{Phi-expl2}
\Phi_{a,b}(\bar u,\bar v)&=&
\sum {K_\ell(\bar v_{\so}|\bar u_{\so})}\, f(\bar u_{\so},\bar u_{\st})\, g(\bar v_{\st},\bar v_{\so})\;
\TT_{13}(\bar v_{\so})\,\TT_{23}(\bar v_{\st})\,T_{12}(\bar u_{\st})\lambda_{2}(\bar u_{\so})\,\Omega.
\eea
where we have introduced the notation (for $\#\bu=\#\bv=\ell$):
\bea\label{TT}
\TT_{23}(\bar v)&=&\frac{1}{H(\bar v)}\,\prod^{\to}_{1\leq j\leq \ell}T_{23}(v_j) \mb{and}
\TT_{13}(\bar v)=\frac{1}{H(\bar v)}\,\prod^{\to}_{1\leq j\leq \ell}T_{13}(v_j),
\\
K_\ell(\bv|\bu) &=&\Delta_\ell(\bv)\,\Delta'_\ell(\bu)\,h(\bv,\bu)\,\det_\ell\left(\frac{g(v_j,u_k)}{h(v_j,u_k)}\right)
\\
\Delta_\ell(\bv)&=&\prod_{\ell\ge j>k\ge 1}g(v_j,v_k)\,,\qquad \Delta'_\ell(\bu)=\prod_{\ell\ge j>k\ge 1}g(u_k,u_j).
\eea

In relations \eqref{Phi-expl} and \eqref{Phi-expl2}, the sum is taken over partitions $\bv\Rightarrow\{\bv_{\so},\bv_{\st}\}$ and $\bu\Rightarrow\{\bu_{\so},\bu_{\st}\}$ with the restriction $\#\bu_{\so}=\#\bv_{\so}=\ell$, $\ell=0,1,..,\min(a,b)$.
\end{prop}
Actually, we will prove a property which is slightly more general. Define
\be{TPhi-expl}
X_{a,b}(\bar u,\bar v)=\sum g(\bv_{\so},\bu_{\so}) f(\bu_{\so},\bu_{\st}) g(\bv_{\st},\bv_{\so})h(\bu_{\so},\bu_{\so})\;
\TT_{13}(\bu_{\so})\,T_{12}(\bu_{\st})\,\TT_{23}(\bv_{\st})\,T_{22}(\bv_{\so}).
\ee
The indices $a,b$ in $X_{a,b}(\bar u,\bar v)$ indicate that $\#\bu=a$ and $\#\bv=b$ and the sum is taken over partitions $\bv\Rightarrow\{\bv_{\so},\bv_{\st}\}$ and $\bu\Rightarrow\{\bu_{\so},\bu_{\st}\}$ with the restriction $\#\bu_{\so}=\#\bv_{\so}$.

We are going to prove that the operator $X_{a,b}(\bar u,\bar v)$ satisfies a recursion
\begin{multline}\label{recurs}
X_{a,b}(\bar u,\bar v) = T_{12}(u_a)\, X_{a-1,b}(\bar u_a,\bar v)\\
+\sum_{j=1}^b g(v_j,u_a)f(v_j,\bar u_a)g(\bv_j,v_j)
\,T_{13}(u_a) X_{a-1,b-1}(\bar u_a,\bar v_j)\,T_{22}(v_j).
\end{multline}
Applying this relation on the highest weight vector $\Omega$ then shows that $X_{a,b}(\bar u,\bar v)\Omega$ obeys the first recursion relation for $\Phi_{ab}(\bar u,\bar v)$. Since they coincide for $ a=0$, it proves that they are equal, so that \eqref{TPhi-expl} provides the explicit expression \eqref{Phi-expl} for $\Phi_{ab}(\bar u,\bar v)$.

In the same way, starting from
\be{TPhi-expl3}
Y_{a,b}(\bu,\bv)=\sum K_{\ell}(\bv_{\so}|\bu_{\so}) f(\bu_{\so},\bu_{\st}) g(\bv_{\st},\bv_{\so})\;
\TT_{13}(\bv_{\so})\,\TT_{23}(\bv_{\st})\,T_{12}(\bu_{\st})T_{22}(\bu_{\so}),
\ee
where the sum is taken over partitions $\bv\Rightarrow\{\bv_{\so},\bv_{\st}\}$ and $\bu\Rightarrow\{\bu_{\so},\bu_{\st}\}$ with the restriction $\#\bu_{\so}=\#\bv_{\so}=\ell$, where $\ell=0,1,\dots, \min(a,b)$, we will show that
 the operator $Y_{a,b}(\bar u,\bar v)$ satisfies a recursion
\begin{multline}\label{recT23}
h(\bv,v_b)Y_{a,b}(\bar u,\bar v)=T_{23}(v_b)Y_{a,b-1}(\bar u,\bar v_b)\\
+\sum_{j=1}^a g(v_b,u_j)f(\bv_b,u_j)f(u_j,\bu_j) T_{13}(v_b) Y_{a-1,b-1}(\bar u_j,\bar v_b)T_{22}(u_j).
\end{multline}
Again, application of this relation on $\Omega$ will show that $Y_{a,b}(\bar u,\bar v)\Omega$ and $\Phi_{a,b}(\bar u,\bar v)$ coincide, and we get the second explicit expression \eqref{Phi-expl2} for $\Phi_{a,b}(\bar u,\bar v)$.

\subsection{Multiple commutation relations\label{S-PMCR}}
Before proving the recursion relations for $X_{a,b}(\bar u,\bar v)$ and $Y_{a,b}(\bar u,\bar v)$, we need some formulas for multiple commutation relations.
\begin{lemma}
Let $\#\bu=\ell$. Then, for $\TT_{13}(\bu)$ defined as in \eqref{TT}, we have
\be{T12T13-1-mp}
T_{12}(v)\TT_{13}(\bu)=f(\bu,v) \TT_{13}(\bu)T_{12}(v) +\sum_{k=1}^\ell g(v,u_k)g(\bu_k,u_k) T_{13}(v)\TT_{13}(\bu_k)T_{12}(u_k),
\ee
\be{Matact2313}
T_{23}(v)\TT_{13}(\bu)=(-1)^\ell f(\bu,v)\TT_{13}(\bu) T_{23}(v)+\sum_{k=1}^\ell g(u_k,v)g(u_k,\bu_k) T_{13}(v)\TT_{13}(\bu_k) T_{23}(u_k) .
\ee
\end{lemma}
\prof
We have from the $RTT$-relation:
\be{T12T13}
T_{12}(v)T_{13}(u)= f(u,v)T_{13}(u)T_{12}(v)+ g(v,u)T_{13}(v)T_{12}(u).
\ee
Then we consider the case of one operator $T_{12}(v)$ and $\ell$ operators $T_{13}(u_k)$. We use the standard approach of the
algebraic Bethe ansatz. It is clear that
\be{T12T13-1-m}
T_{12}(v)\TT_{13}(\bu)=\Lambda \TT_{13}(\bu)T_{12}(v) +\sum_{k=1}^\ell \Lambda_k T_{13}(v)\TT_{13}(\bu_k)T_{12}(u_k),
\ee
where $\Lambda$ and $\Lambda_k$ are some rational coefficients. In order to find $\Lambda$ one should ignore the second
term in the r.h.s. of \eqref{T12T13}. Then
\be{Lam}
\Lambda=f(\bu,v).
\ee

Now let us find $\Lambda_k$. Due to the symmetry of $\TT_{13}(\bu)$ over $\bu$ it is enough to find $\Lambda_1$ only. We have
\bea\label{T12T13-L1}
T_{12}(v)\TT_{13}(\bu)&=&  T_{12}(v)\,\frac{T_{13}(u_1)\dots T_{13}(u_\ell)}{\prod_{\ell\ge j>k\ge 1} h(u_j,u_k)}\\
&=& g(v,u_1)\,\frac{T_{13}(v)T_{12}(u_1)T_{13}(u_2)\dots T_{13}(u_\ell)}{\prod_{\ell\ge j>k\ge 1} h(u_j,u_k)}+UWT,
\eea
where $UWT$ means {\it unwanted terms}. Then the operator $T_{12}(u_1)$ should move to the right keeping its argument,
what gives us
\bea\label{T12T13-L1a}
T_{12}(v)\TT_{13}(\bu)
&=&g(v,u_1)f(\bu_1,u_1)\frac{T_{13}(v)T_{13}(u_2)\dots T_{13}(u_\ell)}{\prod_{\ell\ge j>k\ge 1} h(u_j,u_k)}T_{12}(u_1)+UWT\\
&=&g(v,u_1)g(\bu_1,u_1)T_{13}(v)\TT_{13}(\bu_1)T_{12}(u_1)+UWT.
\eea
Thus, we obtain
\be{Lam1}
\Lambda_1=g(v,u_1)g(\bu_1,u_1),
\ee
and, hence,
\be{Lamk}
\Lambda_k=g(v,u_k)g(\bu_k,u_k).
\ee
Thus, we get \eqref{T12T13-1-mp}.

In the same way, starting from the $RTT$-relation
\be{T23T13}
T_{23}(v)T_{13}(u) =-f(u,v)T_{13}(u)T_{23}(v)-g(v,u) T_{13}(v)T_{23}(u),
\ee
similar considerations show relation \eqref{Matact2313}.
\qed

\subsection{Proof of the recursion for $X_{a,b}(\bar u,\bar v)$\label{S-PR}}

Here we prove that $X_{a,b}(\bar u,\bar v)$ defined by \eqref{TPhi-expl} satisfies the recursion
\eqref{recurs}.

Acting with $T_{12}(u_a)$ onto $X_{a-1,b}(\bu_a;\bv)$ we should move $T_{12}(u_a)$ through the
product $\TT_{13}(\bu_{\so})$. For this we can use \eqref{T12T13-1-m}. It is convenient to rewrite it in the
following form:
\be{T12T13-1-mnew}
T_{12}(v)\TT_{13}(\bu)=f(\bu,v) \TT_{13}(\bu)T_{12}(v) +\sum g(v,u_{\qo})g(\bu_{\qt},u_{\qo}) T_{13}(v)\TT_{13}(\bu_{\qt})T_{12}(u_{\qo}),
\ee
where the sum is taken over partitions $\bu\Rightarrow\{u_{\rm i},\bu_{\rm ii}\}$ with $\# u_{\rm i}=1$ (therefore
we do not write bar for this subset).
Let us call the first term in the r.h.s. of \eqref{T12T13-1-mnew} {\it direct action}, while the remaining
sum over partitions is called  {\it indirect action}.

We have
\be{M1M2}
T_{12}(u_a)X_{a-1,b}(\bu_a;\bv)=M_1 + T_{13}(u_a)M_2,
\ee
where $M_1$ and $M_2$ respectively correspond to the direct and indirect actions of $T_{12}(u_a)$. Consider first the contribution $M_1$.
We have
\be{DA-1}
M_1=\sum g(\bv_{\so},\bu_{\so}) f(\bu_{\so},\bu_{\st}) g(\bv_{\st},\bv_{\so})h(\bu_{\so},\bu_{\so}) f(\bu_{\so},u_a)\;
\TT_{13}(\bu_{\so})\,T_{12}(\bu_{\st})T_{12}(u_a)\,\TT_{23}(\bv_{\st})T_{22}(\bv_{\so}).
\ee
Here  the sum is taken over partitions $\bu_a\Rightarrow\{\bu_{\so},\bu_{\st}\}$ and $\bv\Rightarrow\{\bv_{\so},\bv_{\st}\}$ with
$\#\bv_{\so}=\#\bu_{\so}$. Recall also that  $\bu_a=\bu\setminus \{u_a\}$.
Denoting $\{\bu_{\st},u_a\}=\bu_{\st'}$ we obtain
\be{DA-2}
M_1=\sum_{u_a\in\bu_{\st'}} g(\bv_{\so},\bu_{\so}) f(\bu_{\so},\bu_{\st'}) g(\bv_{\st},\bv_{\so})h(\bu_{\so},\bu_{\so})\;
\TT_{13}(\bu_{\so})\,T_{12}(\bu_{\st'})\,\TT_{23}(\bv_{\st})T_{22}(\bv_{\so}),
\ee
where now the sum is  taken over partitions $\bu\Rightarrow\{\bu_{\so},\bu_{\st'}\}$ and $\bv\Rightarrow\{\bv_{\so},\bv_{\st}\}$ with
$\#\bv_{\so}=\#\bu_{\so}$ and an
additional restriction $u_a\in\bu_{\st'}$. Clearly, if the ignore the latest restriction, then we obtain
$X_{a,b}(\bar u,\bar v)$. Therefore
\be{DA-3}
M_1=X_{a,b}(\bar u,\bar v)-\sum_{u_a\in\bu_{\so}} g(\bv_{\so},\bu_{\so}) f(\bu_{\so},\bu_{\st'}) g(\bv_{\st},\bv_{\so})h(\bu_{\so},\bu_{\so})\;
\TT_{13}(\bu_{\so})\,T_{12}(\bu_{\st})\,\TT_{23}(\bv_{\st'})T_{22}(\bv_{\so}),
\ee
where now the restriction is $u_a\in\bu_{\so}$.
Setting $\bu_{\so}=\{u_a,\bu_{\qt}\}$, we obtain
\begin{multline}\label{DA-4}
M_1-X_{a,b}(\bar u,\bar v)=-\sum g(\bv_{\so},\bu_{\qt})g(\bv_{\so},u_a)  f(u_a,\bu_{\st'})f(\bu_{\qt},\bu_{\st'}) g(\bv_{\st},\bv_{\so})\;
\\
\times h(u_a,\bu_{\qt})h(\bu_{\qt},\bu_{\qt})T_{13}(u_a)\TT_{13}(\bu_{\qt})\,T_{12}(\bu_{\st'})\,\TT_{23}(\bv_{\st})T_{22}(\bv_{\so}),
\end{multline}
where we have extracted explicitly $T_{13}(u_a)$ from the product $\TT_{13}(\bu_{\so})$. Then we recast \eqref{DA-4} as follows
\begin{multline}\label{DA-4a}
M_1-X_{a,b}(\bar u,\bar v)=-T_{13}(u_a)f(u_a,\bu_a)\sum  g(\bv_{\so},\bu_{\qt})\frac{g(\bv_{\so},u_a)}{g(u_a,\bu_{\qt})}f(\bu_{\qt},\bu_{\st'}) g(\bv_{\st},\bv_{\so})h(\bu_{\qt},\bu_{\qt})\;
\\
\times \TT_{13}(\bu_{\qt})\,T_{12}(\bu_{\st'})\,\TT_{23}(\bv_{\st})T_{22}(\bv_{\so}).
\end{multline}
The sum is  taken over partitions $\bu_a\Rightarrow\{\bu_{\qt},\bu_{\st'}\}$ and $\bv\Rightarrow\{\bv_{\so},\bv_{\st}\}$ with
$\#\bv_{\so}=\#\bu_{\qt}+1$.

The final step of the transformations of $M_1$ is to develop
the ratio $g(\bv_{\so},u_a)/g(u_a,\bu_{\qt})$ over the poles at $u_a=v_{\qo}\in\bv_{\so}$:
\be{dev-pol}
\frac{g(\bv_{\so},u_a)}{g(u_a,\bu_{\qt})}=\sum g(v_{\qo},u_a) \frac{g(\bv_{\qt},v_{\qo})}{g(v_{\qo},\bu_{\qt})}.
\ee
Here the sum is taken over partitions $\bv_{\so}\Rightarrow\{v_{\qo},\bv_{\qt}\}$, where $v_{\qo}$ consists of one element. Substituting this into \eqref{DA-4} and setting there $\bv_{\so}=\{v_{\qo},\bv_{\qt}\}$ we obtain
\begin{multline}\label{DA-5}
M_1-X_{a,b}(\bar u,\bar v)
=-T_{13}(u_a)f(u_a,\bu_a)\sum  g(v_{\qo},u_a) g(\bv_{\qt},\bu_{\qt})f(\bu_{\qt},\bu_{\st'})h(\bu_{\qt},\bu_{\qt}) \;
\\
\times g(\bv_{\st},\bv_{\qt})g(\bv_{\st},v_{\qo})
g(\bv_{\qt},v_{\qo})\TT_{13}(\bu_{\qt})\,T_{12}(\bu_{\st'})\,\TT_{23}(\bv_{\st})T_{22}(\bv_{\qt})T_{22}(v_{\qo}).
\end{multline}
This is the final expression for the contribution $M_1$. The sum is organized as follows. The set $\bu_a$ is divided into two
subsets $\bu_a\Rightarrow\{\bu_{\qt},\bu_{\st'}\}$. The set $\bv$ is divided into three subsets $\bv\Rightarrow\{v_{\qo},\bv_{\qt},\bv_{\st}\}$.
The restrictions are  $\#\bv_{\qt}=\#\bu_{\qt}$ and $\#v_{\qo}=1$.

Consider now the result of the indirect action $M_2$. Using \eqref{T12T13-1-mnew} we obtain
\begin{multline}\label{IA-1}
M_2=\sum g(\bv_{\so},\bu_{\so}) f(\bu_{\so},\bu_{\st}) g(\bv_{\st},\bv_{\so})\;
h(\bu_{\so},\bu_{\so})g(u_a,u_{\qo}) g(\bu_{\qt},u_{\qo})\\
\times \TT_{13}(\bu_{\qt})\,T_{12}(\bu_{\st})T_{12}(u_{\qo})\,\TT_{23}(\bv_{\st})T_{22}(\bv_{\so}).
\end{multline}
Here the set $\bu_a$ is divided into $\bu_{a}\Rightarrow\{\bu_{\so},\bu_{\st}\}$, and then the subset $\bu_{\so}$ is divided once more $\bu_{\so}\Rightarrow\{u_{\qo},\bu_{\qt}\}$, where $u_{\qo}$ consists of only one element.

Let $\{\bu_{\qo},\bu_{\st}\}=\bu_{\st'}$. Then we can recast \eqref{IA-1} as follows:
\begin{multline}\label{IA-2}
M_2=\sum \frac{g(\bv_{\so},u_{\qo})}{g(u_{\qo},\bu_{\qt})}g(\bv_{\so},\bu_{\qt})f(u_{\qo},\bu_{0})
 g(u_a,u_{\qo}) f(\bu_{\qt},\bu_{\st'}) g(\bv_{\st},\bv_{\so})h(\bu_{\qt},\bu_{\qt})\\
\times \TT_{13}(\bu_{\qt})\,T_{12}(\bu_{\st'})\,\TT_{23}(\bv_{\st})T_{22}(\bv_{\so}),
\end{multline}
where we also substituted $\bu_{\so}=\{u_{\qo},\bu_{\qt}\}$ and set  $\bu_{0}=\bu_a\setminus \{u_{\qo}\}$.
Now we again develop the ratio $g(\bv_{\so},u_{\qo})/g(u_{\qo},\bu_{\qt})$ over the poles:
\be{rat-poles}
\frac{g(\bv_{\so},u_{\qo})}{g(u_{\qo},\bu_{\qt})}=\sum g(v_{\qo},u_{\qo})\frac{g(\bv_{\qt},v_{\qo})}{g(v_{\qo},\bu_{\qt})},
\ee
where the sum is taken over partitions $\bv_{\so}\Rightarrow\{v_{\qo},\bv_{\qt}\}$, and $v_{\qo}$ consists of one element only. Then
\eqref{IA-2} takes the form
\begin{multline}\label{IA-3}
M_2=\sum \Bigl[g(u_a,u_{\qo})g(v_{\qo},u_{\qo})f(u_{\qo},\bu_{0})\Bigr]
g(\bv_{\qt},\bu_{\qt})
  f(\bu_{\qt},\bu_{\st'})h(\bu_{\qt},\bu_{\qt})\\
\times g(\bv_{\qt},v_{\qo})g(\bv_{\st},v_{\qo}) g(\bv_{\st},\bv_{\qt}) \TT_{13}(\bu_{\qt})\,T_{12}(\bu_{\st'})\,\TT_{23}(\bv_{\st})T_{22}(\bv_{\qt})T_{22}(v_{\qo}).
\end{multline}
Here the sum is taken over partitions $\bu_a\Rightarrow\{u_{\qo},\bu_{\qt},\bu_{\st'}\}$ and $\bv\Rightarrow\{v_{\qo},\bv_{\qt},\bv_{\st}\}$
with the restrictions $\#\bv_{\qt}=\#\bu_{\qt}$ and $\#v_{\qo}=\#u_{\qo}=1$.
The sum over $u_{\qo}$ (see the terms in the square brackets in \eqref{IA-3}) can be computed via a special contour integral. Let
\be{sum-ui0}
\sum_{u_{\qo}\in\bu_a}f(u_{\qo},\bu_{0})g(u_a,u_{\qo})g(v_{\qo},u_{\qo})=J.
\ee
Consider an integral
\be{int-sum}
I=\frac{c}{2\pi i}\oint_{|z|=R\to\infty}\frac{dz}{(u_a-z)(v_{\qo}-z)}\prod_{k=1}^{a-1}\frac{z-u_k+c}{z-u_k},
\ee
where the integration contour is $|z|=R\to\infty$. Obviously, $I=0$, because the integrand behaves as $z^{-2}$ at $|z|\to\infty$.
On the other hand, this integral is equal to the sum of residues within the integration contour. The sum of the residues in the
poles $z=u_k$ gives $J$. Two additional contributions come from the poles at $z=u_a$ and $z=v_{\qo}$. Thus, we obtain
\be{sum-ui}
0=J-g(v_{\qo},u_a)f(u_a,\bu_a)-g(v_{\qo},u_a)f(v_{\qo},\bu_a).
\ee
Substituting this into \eqref{IA-3} we find
\begin{multline}\label{IA-4}
M_2=\sum g(v_{\qo},u_a)\Bigl\{f(u_a,\bu_a)-f(v_{\qo},\bu_a)\Bigr\}
g(\bv_{\qt},\bu_{\qt})
  f(\bu_{\qt},\bu_{\st'})h(\bu_{\qt},\bu_{\qt}) \\
\times g(\bv_{\qt},v_{\qo})g(\bv_{\st},v_{\qo}) g(\bv_{\st},\bv_{\qt})
\TT_{13}(\bu_{\qt})\,T_{12}(\bu_{\st'})\,\TT_{23}(\bv_{\st})T_{22}(\bv_{\qt})T_{22}(v_{\qo}).
\end{multline}
We see that the first term in the braces cancels the contribution \eqref{DA-5}. Thus, we arrive at
\begin{multline}\label{DAIA-1}
T_{12}(u_a)X_{a-1,b}(\bar u_a,\bar v)-X_{a,b}(\bar u,\bar v)=
-T_{13}(u_a)\sum g(\bv_{\qt},\bu_{\qt}) g(v_{\qo},u_a)f(v_{\qo},\bu_a)f(\bu_{\qt},\bu_{\st'})h(\bu_{\qt},\bu_{\qt})\\
\times
    g(\bv_{\qt},v_{\qo})g(\bv_{\st},v_{\qo})g(\bv_{\st},\bv_{\qt}) \;
 \TT_{13}(\bu_{\qt})\,T_{12}(\bu_{\st'})\,\TT_{23}(\bv_{\st})T_{22}(\bv_{\qt})T_{22}(v_{\qo}).
\end{multline}
Here the sum is taken over partitions $\bu_a\Rightarrow\{\bu_{\qt},\bu_{\st'}\}$ and $\bv\Rightarrow\{v_{\qo},\bv_{\qt},\bv_{\st}\}$
with the restrictions $\#\bv_{\qt}=\#\bu_{\qt}$ and $\#v_{\qo}=1$. Let $\bv_0=\bv\setminus \{v_{\qo}\}=\{\bv_{\qt},\bv_{\st}\}$. Then
\eqref{DAIA-1} takes the form
\begin{multline}\label{DAIA-1a}
T_{12}(u_a)X_{a-1,b}(\bar u_a,\bar v)-X_{a,b}(\bar u,\bar v)=
-T_{13}(u_a)\sum g(v_{\qo},u_a)f(v_{\qo},\bu_a)g(\bv_{0},v_{\qo})\\
\times \Bigl\{g(\bv_{\qt},\bu_{\qt})
   f(\bu_{\qt},\bu_{\st'}) g(\bv_{\st},\bv_{\qt})h(\bu_{\qt},\bu_{\qt}) \;
 \TT_{13}(\bu_{\qt})\,T_{12}(\bu_{\st'})\,\TT_{23}(\bv_{\st})T_{22}(\bv_{\qt})\Bigr\}T_{22}(v_{\qo}).
\end{multline}
The sum over partitions in the braces evidently gives $X_{a-1,b-1}(\bar u_a,\bar v_{\qo})$, and we finally obtain
\begin{equation}\label{DAIA-2}
T_{12}(u_a)X_{a-1,b}(\bar u_a,\bar v)
=X_{a,b}(\bar u,\bar v)-\sum g(v_{\qo},u_a)f(v_{\qo},\bu_a)g(\bv_{0},v_{\qo})T_{13}(u_a)X_{a-1,b-1}(\bar u_a,\bar v_{\qo})T_{22}(v_{\qo}).
\end{equation}
This is exactly the recursion that we need.


\subsection{Proof of the recursion for $Y_{a,b}(\bar u,\bar v)$}

The proof if very similar to the one given in section~\ref{S-PR}.
Now we should multiply the operator $Y_{a,b-1}(\bar u,\bar v_b)$ by  $T_{23}(v_b)$ from the left and move  $T_{23}(v_b)$
through the product $\TT_{13}(\bv_{\so})$. The result can be written
as a sum of two terms:
\be{act-MM}
T_{23}(v_b)Y_{a,b-1}(\bar u,\bar v_b)=M_1+ T_{13}(v_b)M_2.
\ee
Here the contributions $M_1$ and $M_2$ respectively correspond to the first and the second terms in the action \eqref{Matact2313}. Consider the first
contribution
\be{M1-1}
M_1=\sum (-1)^\ell K_\ell(\bv_{\so}|\bu_{\so}) f(\bu_{\so},\bu_{\st}) g(\bv_{\st},\bv_{\so})f(\bv_{\so},v_b)\;
\TT_{13}(\bv_{\so})\,T_{23}(v_b)\TT_{23}(\bv_{\st})\,T_{12}(\bu_{\st})T_{22}(\bu_{\so}).
\ee
Here the sum is taken over partitions $\bv_b\Rightarrow\{\bv_{\so},\bv_{\st}\}$ and $\bu\Rightarrow\{\bu_{\so},\bu_{\st}\}$ with the restriction $\#\bu_{\so}=\#\bv_{\so}=\ell$.   Combining $\{v_b,\bv_{\st}\}=\bv_{\st'}$, we obtain
\be{M1-2}
M_1=h(\bv_b,v_b)\sum_{v_b\in\bv_{\st'}} K_\ell(\bv_{\so}|\bu_{\so}) f(\bu_{\so},\bu_{\st}) g(\bv_{\st'},\bv_{\so})\;
\TT_{13}(\bv_{\so})\,\TT_{23}(\bv_{\st'})\,T_{12}(\bu_{\st})T_{22}(\bu_{\so}),
\ee
where now the sum is taken over partitions of the complete set $\bv$ into subsets $\bv_{\so}$ and $\bv_{\st'}$. However, we have
the restriction $v_b\in\bv_{\st'}$. Obviously, if we get rid of this restriction, then
we obtain $h(\bv_b,v_b)Y_{a,b}(\bu,\bv)$. Thus,
\begin{multline}\label{M1-3}
M_1-h(\bv_b,v_b)Y_{a,b}(\bu,\bv)\\
=-h(\bv_b,v_b)\sum_{v_b\in\bv_{\so}} K_\ell(\bv_{\so}|\bu_{\so}) f(\bu_{\so},\bu_{\st}) g(\bv_{\st'},\bv_{\so})\;
\TT_{13}(\bv_{\so})\,\TT_{23}(\bv_{\st'})\,T_{12}(\bu_{\st})T_{22}(\bu_{\so}).
\end{multline}
Here the sum over partitions is the same as in \eqref{M1-2} except that now
the restriction is $v_b\in\bv_{\so}$. Therefore we can set $\bv_{\so}=\{\bv_{\qt},v_b\}$ and
develop $K_\ell(\bv_{\so}|\bu_{\so})$ over the residues at $v_b=u_{\qo}$, where $u_{\qo}\in\bu_{\so}$.
Let  $\bu_{\qt}=\bu_{\so}\setminus \{u_{\qo}\}$. Then
\be{dev-K}
K_\ell(\bv_{\so}|\bu_{\so})=\sum g(v_b,u_{\qo})f(\bv_{\qt},u_{\qo})f(u_{\qo},\bu_{\qt})K_{\ell-1}(\bv_{\qt}|\bu_{\qt}),
\ee
where the sum is taken over partitions $\bu_{\so}\Rightarrow\{\bu_{\qt},u_{\qo}\}$, and the subset $u_{\qo}$ consists of one element.
Substituting this into \eqref{M1-3} we arrive at
\begin{multline}\label{M1-4}
M_1-h(\bv_b,v_b)Y_{a,b}(\bu,\bv)
=-h(\bv_b,v_b)\sum g(v_b,u_{\qo})f(\bv_{\qt},u_{\qo})f(u_{\qo},\bu_{\qt})K_{\ell-1}(\bv_{\qt}|\bu_{\qt}) \\
\times f(\bu_{\qt},\bu_{\st})f(u_{\qo},\bu_{\st}) g(\bv_{\st'},\bv_{\qt})g(\bv_{\st'},v_b)\;
\frac{T_{13}(v_b)\TT_{13}(\bv_{\qt})}{h(\bv_{\qt},v_b)}\TT_{23}(\bv_{\st'})\,T_{12}(\bu_{\st})T_{22}(\bu_{\qt})T_{22}(u_{\qo}).
\end{multline}
Here the sum is taken over partitions $\bv_b\Rightarrow\{\bv_{\qt},\bv_{\st'}\}$ and $\bu\Rightarrow\{u_{\qo},\bu_{\qt},\bu_{\st}\}$ with the restrictions $\#\bu_{\qt}=\#\bv_{\qt}=\ell-1$ and $\#u_{\qo}=1$.
Setting here $\bu_{0}=\bu\setminus \{u_{\qo}\}$ we finally obtain
\begin{multline}\label{M1-5}
M_1-h(\bv_b,v_b)Y_{a,b}(\bu,\bv)
=-T_{13}(v_b)\sum g(v_b,u_{\qo})f(\bv_{\qt},u_{\qo})f(\bv_{\st'},v_b)f(u_{\qo},\bu_{0}) \\
\times K_{\ell-1}(\bv_{\qt}|\bu_{\qt})f(\bu_{\qt},\bu_{\st}) g(\bv_{\st'},\bv_{\qt})\;
\TT_{13}(\bv_{\qt})\TT_{23}(\bv_{\st'})\,T_{12}(\bu_{\st})T_{22}(\bu_{\qt})T_{22}(u_{\qo}).
\end{multline}

Let us consider now the contribution $M_2$:
\begin{equation}\label{M2-1}
M_2=\sum K_\ell(\bv_{\so}|\bu_{\so}) f(\bu_{\so},\bu_{\st}) g(\bv_{\st},\bv_{\qt})g(v_{\qo},v_b)g(v_{\qo},\bv_{\qt})
%
\TT_{13}(\bv_{\qt})\,T_{23}(v_{\qo})\TT_{23}(\bv_{\st})\,T_{12}(\bu_{\st})T_{22}(\bu_{\so}),
\end{equation}
Here the sum is taken over partitions $\bv_b\Rightarrow\{\bv_{\so},\bv_{\st}\}$ and $\bu\Rightarrow\{\bu_{\so},\bu_{\st}\}$ with the restriction $\#\bu_{\so}=\#\bv_{\so}=\ell$, and then the subset $\bv_{\so}$ is divided into $v_{\qo}$ and $\bv_{\qt}$, where $v_{\qo}$ consists of one element.
The goal is to combine $v_{\qo}$ and $\bv_{\st}$ into subset $\bv_{\st'}$. For this we first transform
\eqref{M2-1} as follows:
\begin{equation}\label{M2-2}
M_2=\sum (-1)^\ell K_\ell(\{\bv_{\qt},v_{\qo}\}|\bu_{\so})g(\bv_0,v_{\qo}) f(\bu_{\so},\bu_{\st}) g(\bv_{\st},\bv_{\qt})
%
\TT_{13}(\bv_{\qt})\,T_{23}(v_{\qo})\TT_{23}(\bv_{\st})\,T_{12}(\bu_{\st})T_{22}(\bu_{\so}),
\end{equation}
where $\bv_0=\bv\setminus \{v_{\qo}\}$. Then we introduce $\bv_{\st'}=\{v_{\qo},\bv_{\st}\}$ and obtain
\begin{multline}\label{M2-3}
M_2=\sum (-1)^\ell K_\ell(\{\bv_{\qt},v_{\qo}\}|\bu_{\so})\frac{g(\bv_0,v_{\qo})}{g(v_{\qo},\bv_{\qt})} f(\bu_{\so},\bu_{\st}) g(\bv_{\st'},\bv_{\qt})
 h(\bv_{\st'},v_{\qo})\\
\times \TT_{13}(\bv_{\qt})\,\TT_{23}(\bv_{\st'})\,T_{12}(\bu_{\st})T_{22}(\bu_{\so}).
\end{multline}
Now we again develop $K_\ell(\{\bv_{\qt},v_{\qo}\}|\bu_{\so})$ with respect to the poles at $v_{\qo}=u_{\qo}$, $u_{\qo}\in\bu_{\so}$. This gives us
\begin{multline}\label{M2-4}
M_2=-\sum  \Bigl[g(v_{\qo},u_{\qo})g(v_b,v_{\qo})f(\bv_{\st},v_{\qo})\Bigr]f(\bv_{\qt},u_{\qo})f(u_{\qo},\bu_{0}) \\
\times K_{\ell-1}(\bv_{\qt}|\bu_{\qt}) f(\bu_{\qt},\bu_{\st}) g(\bv_{\st'},\bv_{\qt})
\TT_{13}(\bv_{\qt})\,\TT_{23}(\bv_{\st'})\,T_{12}(\bu_{\st})T_{22}(\bu_{\qt})T_{22}(u_{\qo}).
\end{multline}
Here the sum is taken over partitions $\bv_b\Rightarrow\{v_{\qo},\bv_{\qt},\bv_{\st}\}$ and $\bu\Rightarrow\{u_{\qo},\bu_{\qt},\bu_{\st}\}$ with the restrictions $\#\bu_{\qt}=\#\bv_{\qt}=\ell-1$ and $\# u_{\qo}=\# v_{\qo}=1$. Hereby $\bv_{\st'}=\{v_{\qo},\bv_{\st}\}$ and we introduced
 $\bu_{0}=\bu\setminus \{u_{\qo}\}$.

Now  we can take the sum over $v_{\qo}$ (see the terms in the square brackets in \eqref{M2-4}). Recall that  $v_{\qo}$ runs through the subset $\bv_{\st'}$. Therefore, we easily find by means of
contour integration
\be{sum-v0}
\sum_{v_{\qo}\in \bv_{\st'}}g(v_{\qo},u_{\qo})g(v_b,v_{\qo})f(\bv_{\st},v_{\qo})=g(v_b,u_{\qo})\Bigl\{f(\bv_{\st'},u_{\qo})-f(\bv_{\st'},v_b)\Bigr\}.
\ee
Hence,
\begin{multline}\label{M2-6}
M_2=-\sum f(\bv_{\qt},u_{\qo})f(u_{\qo},\bu_{0})g(v_b,u_{\qo})\Bigl\{f(\bv_{\st'},u_{\qo})-f(\bv_{\st'},v_b)\Bigr\} \\
\times K_{\ell-1}(\bv_{\qt}|\bu_{\qt}) f(\bu_{\qt},\bu_{\st}) g(\bv_{\st'},\bv_{\qt})
\TT_{13}(\bv_{\qt})\,\TT_{23}(\bv_{\st'})\,T_{12}(\bu_{\st})T_{22}(\bu_{\qt})T_{22}(u_{\qo}) .
\end{multline}
Finally, combining \eqref{M2-6} and  \eqref{M1-5} we arrive at
\begin{multline}\label{M12-fin}
T_{23}(v_b)Y_{a,b-1}(\bu,\bv_b)-h(\bv_b,v_b)Y_{a,b}(\bu,\bv)
=-T_{13}(v_b)\sum  g(v_b,u_{\qo})f(\bv_b,u_{\qo})f(u_{\qo},\bu_{0})\\
\times \Bigl\{K_{\ell-1}(\bv_{\qt}|\bu_{\qt}) f(\bu_{\qt},\bu_{\st}) g(\bv_{\st'},\bv_{\qt})
\TT_{13}(\bv_{\qt})\,\TT_{23}(\bv_{\st'})\,T_{12}(\bu_{\st})T_{22}(\bu_{\qt})\Bigr\}T_{22}(u_{\qo}).
\end{multline}
The sum in the r.h.s. is organized as follows. First we divide the set $\bu$ into subset $u_{\qo}$ (with $\# u_{\qo}=1$)
and the complementary subset $\bu_{0}$. Then the have additional partitions $\bu_{0}\Rightarrow\{\bu_{\qt},\bu_{\st}\}$
and $\bv_{b}\Rightarrow\{\bv_{\qt},\bv_{\st'}\}$ with $\#\bu_{\qt}=\#\bv_{\qt}=\ell-1$ (see the terms in the braces in the
second line of \eqref{M12-fin}). Clearly the sum over these partitions gives $Y_{a-1,b-1}(\bu_{0},\bv_b)$, and we arrive at
\begin{multline}\label{M12-fin1}
T_{23}(v_b)Y_{a,b-1}(\bu,\bv_b)-h(\bv_b,v_b)Y_{a,b}(\bu,\bv)\\
=-\sum  g(v_b,u_{\qo})f(\bv_b,u_{\qo})f(u_{\qo},\bu_{0})T_{13}(v_b) Y_{a-1,b-1}(\bu_{0},\bv_b)T_{22}(u_{\qo}),
\end{multline}
which gives us the recursion \eqref{recT23}.

\section{Expressions for $Y(1|2)$ Bethe vectors and dual Bethe vectors\label{sect:expl-remain}}
Once we have explicit expressions for Bethe vectors in $Y(2|1)$, the morphisms $\vph$, $\psi$ and $\wt\psi$ allow us to get  explicit expressions for the remaining (dual) Bethe vectors, as detailed in the following propositions.
\begin{prop}
In $Y(1|2)$ the Bethe vectors have the two following explicit expressions
\bea
\wt\Phi_{a,b}(\bar u,\bar v)&=&
\sum (-1)^{b}\,g(\bar u_{\so},\bar v_{\so})\, f(\bar v_{\so},\bar v_{\st}) \,g(\bar u_{\st},\bar u_{\so})\, {h(\bar v_{\so},\bar v_{\so})}\;
\wt\TT_{13}(\bar v_{\so})\,\wt T_{23}(\bar v_{\st})\,\wt\TT_{12}(\bar u_{\st})\,\wt\lambda_2(\bar u_{\so})\,\wt\Omega,
\nonu
\label{wtPhi-expl}\\
\label{wtPhi-expl2}
\wt\Phi_{a,b}(\bar u,\bar v)&=&
\sum (-1)^{b}\,{K_\ell(\bar u_{\so}|\bar v_{\so})}\, f(\bar v_{\so},\bar v_{\st})\, g(\bar u_{\st},\bar u_{\so})\;
\wt\TT_{13}(\bar u_{\so})\,\wt\TT_{12}(\bar u_{\st})\,\wt T_{23}(\bar v_{\st})\,\wt\lambda_{2}(\bar v_{\so})\,\wt\Omega.
\eea
In relations \eqref{wtPhi-expl} and \eqref{wtPhi-expl2}, the sum is taken over partitions $\bv\Rightarrow\{\bv_{\so},\bv_{\st}\}$ and $\bu\Rightarrow\{\bu_{\so},\bu_{\st}\}$ with the restriction $\#\bu_{\so}=\#\bv_{\so}=\ell$, $0\leq \ell\leq \min(a,b)$. We have also introduced
 (for $\#\bar z=k$)
\be{wtTT}
\wt\TT_{12}(\bar z)=\frac{1}{H(\bar z)}\,\prod^{\to}_{1\leq j\leq k}\wt T_{12}(z_j) \mb{and}
\wt\TT_{13}(\bar z)=\frac{1}{H(\bar z)}\,\prod^{\to}_{1\leq j\leq k}\wt T_{13}(z_j).
\ee
\end{prop}
\prof
We obtain the relations by application of $\vph$ and $\wt\vph$ to the relations \eqref{Phi-expl} and \eqref{Phi-expl2}. The proof is similar to  the ones of section \ref{sect:2ndrecur}.
\qed

\begin{prop}
In $Y(2|1)$, the dual Bethe vectors $\Psi_{ab}(\bar u,\bar v)$ comply the following explicit expressions:
\bea
\Psi_{a,b}(\bar u,\bar v) &=&\!
\Omega^\dag\,(-1)^{(b-1)b/2}\sum g(\bar v_{\so},\bar u_{\so})\, f(\bar u_{\so},\bar u_{\st})\, g(\bar v_{\st},\bar v_{\so})\,
{h(\bar u_{\so},\bar u_{\so})}\,
\lambda_2(\bar v_{\so})\,
\TT_{32}(\bar v_{\st})\,T_{21}(\bar u_{\st})\,\TT_{31}(\bar u_{\so}),
\nonu
\label{dual1}\\
\Psi_{a,b}(\bar u,\bar v) &=&
\Omega^\dag\,(-1)^{(b-1)b/2}\sum
{K_\ell(\bar v_{\so}|\bar u_{\so})}\, f(\bar u_{\so},\bar u_{\st}) \,g(\bar v_{\st},\bar v_{\so})\,\lambda_{2}(\bar u_{\so})\,
T_{21}(\bar u_{\st})\,\TT_{32}(\bar v_{\st})\,\TT_{31}(\bar v_{\so}),
\label{dual2}
\eea
In $Y(1|2)$, the dual Bethe vectors $\wt\Psi_{ab}(\bar u,\bar v)$ have the following explicit expressions:
\bea
\wt\Psi_{a,b}(\bar u,\bar v) &=&
\wt\Omega^\dag\,(-1)^{b+(a-1)a/2}\sum 
g(\bar u_{\so},\bar v_{\so})\, f(\bar v_{\so},\bar v_{\st}) \,g(\bar u_{\st},\bar u_{\so})\, {h(\bar v_{\so},\bar v_{\so})}\;
\wt\lambda_2(\bar u_{\so})
\nonu
&&\qquad\qquad\times\,\wt\TT_{21}(\bar u_{\st})\,\wt T_{32}(\bar v_{\st})\,\wt\TT_{31}(\bar v_{\so})
\label{wtdual1}\\
\wt\Psi_{a,b}(\bar u,\bar v) &=&
\wt\Omega^\dag\,(-1)^{b+(a-1)a/2}\sum \,
{K_\ell(\bar u_{\so}|\bar v_{\so})}\, f(\bar v_{\so},\bar v_{\st})\, g(\bar u_{\st},\bar u_{\so})\,\wt\lambda_{2}(\bar v_{\so})\;
\nonu
&&\qquad\qquad\times\,\wt T_{32}(\bar v_{\st})\,\wt\TT_{21}(\bar u_{\st})\,\wt\TT_{31}(\bar u_{\so})\,.
\label{wtdual2}
\eea
Again, the sums are taken over partitions $\bv\Rightarrow\{\bv_{\so},\bv_{\st}\}$ and $\bu\Rightarrow\{\bu_{\so},\bu_{\st}\}$ with the restriction $0\leq \#\bu_{\so}=\#\bv_{\so}=\ell\leq \min(a,b)$. We have introduced
\bea
&&\TT_{32}(\bar v)=\frac{1}{H(\bar v^*)}\,\prod^{\to}_{1\leq j\leq \ell}T_{32}(v_j) \mb{,}
\TT_{31}(\bar v)=\frac{1}{H(\bar v^*)}\,\prod^{\to}_{1\leq j\leq \ell}T_{31}(v_j),
\\
&&
\wt\TT_{21}(\bar u)=\frac{1}{H(\bar u^*)}\,\prod^{\to}_{1\leq j\leq \ell}\wt T_{21}(u_j)
\mb{and} \wt\TT_{31}(\bar u)=\frac{1}{H(\bar u^*)}\,\prod^{\to}_{1\leq j\leq \ell}\wt T_{31}(u_j).
\eea
\end{prop}
\prof
To get \eqref{dual1} and \eqref{dual2}, we apply the antimorphism $\psi$ to the explicit expressions of BVs \eqref{Phi-expl} and \eqref{Phi-expl2}. Expressions \eqref{wtdual1} and \eqref{wtdual2} are obtained from \eqref{wtPhi-expl} and \eqref{wtPhi-expl2} with the use of $\wt\psi$. The proof is similar to the ones of section \ref{sect:dual}. In particular, it makes appear $H(\bar v^*)$ for the definition of
e.g. $\TT_{32}(\bar v)$, as in the super-trace formula for dual Bethe vectors.
\qed

\section*{Acknowledgements}

 Work of S.P. was supported in part by RFBR grant 16-01-00562. 
N.A.S. was  supported by grants RFBR-15-31-20484-mol-a-ved, RFBR-14-01-00860-a.

\end{document}